\documentclass[twocolumn]{aastex63}
\usepackage{amsmath}

\newcommand{\blue}[1]{{#1}}

\shorttitle{Remote sensing of prominence forces}
\shortauthors{Uritsky, Thompson, and DeVore}

\begin{document}

\title{Remote Sensing of Coronal Forces During a Solar Prominence Eruption}

\correspondingauthor{Vadim Uritsky}
\email{vadim.uritsky@nasa.gov}

\author[0000-0000-0000-0000]{V.\ M.\ Uritsky}
\affiliation{Catholic University of America, 620 Michigan Avenue NE, Washington DC 20061, USA}
\affiliation{Heliophysics Science Division, NASA Goddard Space Flight Center, 8800 Greenbelt Road,
Greenbelt MD 20771, USA}

\author[0000-0000-0000-0000]{B.\ J.\ Thompson}
\affiliation{Heliophysics Science Division, NASA Goddard Space Flight Center, 8800 Greenbelt Road,
Greenbelt MD 20771, USA}

\author[0000-0000-0000-0000]{C.\ R.\ DeVore}
\affiliation{Heliophysics Science Division, NASA Goddard Space Flight Center, 8800 Greenbelt Road,
Greenbelt MD 20771, USA}

\begin{abstract}

We present a new methodology -- the Keplerian Optical Dynamics Analysis (KODA) -- for analyzing the dynamics of dense, cool material in the solar corona. The technique involves adaptive spatiotemporal tracking of propagating intensity gradients and their characterization in terms of time-evolving Keplerian areas swept out by the position vectors of moving plasma blobs. 
Whereas gravity induces purely ballistic motions consistent with Kepler's second law, non-central forces such as the Lorentz force introduce non-zero torques resulting in more complex motions. KODA algorithms enable direct evaluation of the line-of-sight component of the net torque density from the image-plane projection of the areal acceleration. The method is applied to the prominence eruption of 2011 June 7, observed by the Solar Dynamics Observatory's Atmospheric Imaging Assembly. Results obtained include quantitative estimates of the magnetic forces, field intensities, and blob masses and energies across a vast region impacted by the post-reconnection redistribution of the prominence material. The magnetic pressure and energy are strongly dominant during the early, rising phase of the eruption, while the dynamic pressure and kinetic energy become significant contributors during the subsequent falling phases. Measured intensive properties of the prominence blobs are consistent with those of typical active-region prominences; measured extensive properties are compared with those of the whole pre-eruption prominence and the post-eruption coronal mass ejection of 2011 June 7, all derived by other investigators and techniques. The results suggest that the developed technique provides valuable information on characteristics of erupting prominences that are not readily available via alternative means, thereby shedding new light on the environment and evolution of these violent solar events.

\end{abstract}


\section{Introduction} \label{sec:intro}

Eruptions from the Sun of giant prominences (at the limb) and filaments (on the disk) are among the most spectacular and energetic transient events in the solar system. The prominences/filaments are the largest coherent structures in the solar atmosphere; their lengths can rival the visible radius of the star in the most impressive instances. Within their large-scale superstructure, they exhibit fine-scale threads and knots near the limits of observational resolution, and continuous dynamics of counterstreaming quasi-horizontal flows (in many prominences) and/or circulatory quasi-vertical flows (specifically in ``hedgerow'' prominences). Moreover, the threads and knots consist of plasma that is approximately two orders of magnitude denser and cooler than the plasma of the enveloping corona. The myriad challenges of understanding the structure and dynamics of the magnetic field and plasma of these objects across different scales and solar cycle phases have captured the attention of heliophysicists for well over a century \citep[for some reviews, see][]{muzalevskii70, tandberg-hanssen74,tandberg-hanssen95,priest89,labrosse10,mackay10,parenti14,vial15,gibson18}.

In this paper, we present a new image processing pipeline, the Keplerian Optical Dynamics Analysis (KODA), for remotely sensing physical properties of an erupting prominence through kinematic analyses of its rising and falling material. The key underlying idea is to track fragments of the prominence plasma as ``test particles'' whose non-ballistic motion communicates information about the coronal environment. The methodology is applied to the well-studied 2011 June 7 prominence eruption, though it is expected to be applicable to other events observed under similar conditions. We relate our results and conclusions to those of many other investigations of the 2011 June 7 eruptive flare and coronal mass ejection. 

The analytical approach underlying KODA algorithms includes three main processing steps. (1) Adaptive spatiotemporal tracking of moving pieces of fragmented prominence material, referred to as ``blobs'' in the following text, enables the identification of non-ballistic trajectory perturbations caused by the interaction of the prominence plasma with ambient coronal field. (2) The kinematic, dynamic, and energetic parameters characterizing each successfully tracked trajectory are evaluated. (3) The large-scale coronal properties of the eruption are reconstructed by ensemble-averaging over the numerous prominence blobs detected at different stages. Output parameters include the mass of the returning prominence material, the strength of the magnetic force acting on the prominence blobs, the average magnetic field strength in the corona, the energy partitioning in the erupted matter, and other relevant characteristics. 

The paper is organized as follows. \S 2 provides a description of the studied prominence eruption event and the data used in this study. \S 3 presents methods and algorithms involved in KODA. \S 4 presents the results of our investigation of the 2011 June 7 eruptive event. \S 5 discusses our results in the context of earlier studies. \S 6 summarizes our findings and outlines possible future applications of the presented methodology. 


\section{2011 June 7 Prominence Eruption}
\label{sec:event}

On 2011 June 7 a spectacular prominence eruption from NOAA Active Region (AR) 11226, accompanied by a coronal mass ejection (CME) and an M2.5-class eruptive flare, was observed by multiple instruments. 
Photospheric magnetic-field measurements in the days prior to eruption revealed substantial flux cancellation  occurring in the vicinity of the prominence \citep{yardley16}. 
As the eruptive event got underway, the magnetic field of the prominence slowly assumed an increasingly vertical orientation during the initial slow rise, then transitioned rapidly toward a more horizontal orientation after onset of the impulsive upward acceleration \citep{fainshtein16,fainshtein17,egorov20}. 
Strong flare emission was detected during the interval 06:20-06:46 UT in the extreme ultraviolet (EUV) and hard X-rays (HXR), as well as in $\gamma$-rays \citep{inglis13,ackermann14}. 
The violent eruption also launched a rapidly accelerating, quickly moving EUV wave \citep{cheng12,li12} that traversed the corona, generated type-II and type-IV radio bursts and other microwave signatures \citep{cheng12,katoh14,dorovskyy15,susino15,karlicky20}, and drove a white-light shock ahead of the CME into the inner heliosphere \citep{susino15,wood16}. 

Interactions of the erupting prominence with overlying and neighboring magnetic flux produced a fountain-like spray of fragmented cool, dense material that mostly fell back to the solar surface into regions surrounding the initial prominence location \citep{thompson16}. 
Observations of the photospheric field and the coronal plasma evolution, together with magnetohydrodynamics (MHD) modeling, were used to interpret changes in the local connectivity of the coronal field due to magnetic reconnection and the plasma dynamics observed during the event \citep{driel14,petralia16,dudik19}. 
It has been conjectured that the fingering and fragmentation of the erupted prominence plasma were due to onset of Rayleigh-Taylor instability \citep{innes12,carlyle14,mishra18}. 
The density, temperature, and kinetic energy content of the plasma blobs were estimated using multispectral EUV observations during transit through the corona and upon impact with the chromosphere below \citep{gilbert13,landi13,reale14,innes16}. 
A portion of the ejected prominence left the Sun with the CME and was tracked all the way to 1 AU \citep{wood16}. 

For this study, we analyzed a sequence of 193 {\AA} solar coronal images obtained from the Atmospheric Imaging Assembly \citep[AIA;][]{lemen12} onboard the Solar Dynamics Observatory \citep[SDO;][]{pesnell12} spacecraft covering the time interval between 05:40 UT and 08:00 UT on 2011 June 7. SDO/AIA returns images of the full solar disk with the spatial resolution of about 0.6 arcsec. The cool prominence material is observed as dark structures in many AIA channels including those at 131, 171, 193, 211, 304, and 335 {\AA}. The dark structures are caused by the absorption of the background emission by neutral hydrogen and helium atoms and by singly ionized helium ions in the cool plasma. To track the falling prominence material, we opted to use 193 {\AA} SDO/AIA images due to their high contrast and relatively low signal-to-noise ratio, which enabled us to obtain well-defined determinations of the shape and position of the falling prominence pieces. 

The M2.5 flare that occurred during the studied time period as part of the eruption was of sufficient magnitude to trigger an onboard flare observation mode from 06:22 UT to 07:57 UT, in which the AIA instrument alternates between normal 2-s exposures and adaptive exposures as short as 0.12 s.  The images were collected at a 12-s cadence, with a normal 2-s exposure every 24 s and the reduced exposures between them.  For many purposes, the standard exposure time normalization is sufficient to mitigate the difference between the average photon fluxes detected at the two integration times, and obtain a uniform image sequence at the nominal 12-s cadence. However, we found this approach to be inconsistent with our image processing methodology. The exposure time normalization leads to a small but systematic oscillation of the average photon flux in the dark features over the 24-s time scale, which interferes with the threshold-based detection of some moving features and disrupts their temporal tracking history. To exclude this artifact, we studied only the AIA images with the standard exposure time and excluded the shorter exposure images. Reducing to the 24-s cadence had little to no impact on the measured parameters, as the time scale of the changes we measured was substantially longer than 24 s.

\begin{figure*}
\begin{center}
\includegraphics[width=16. cm]{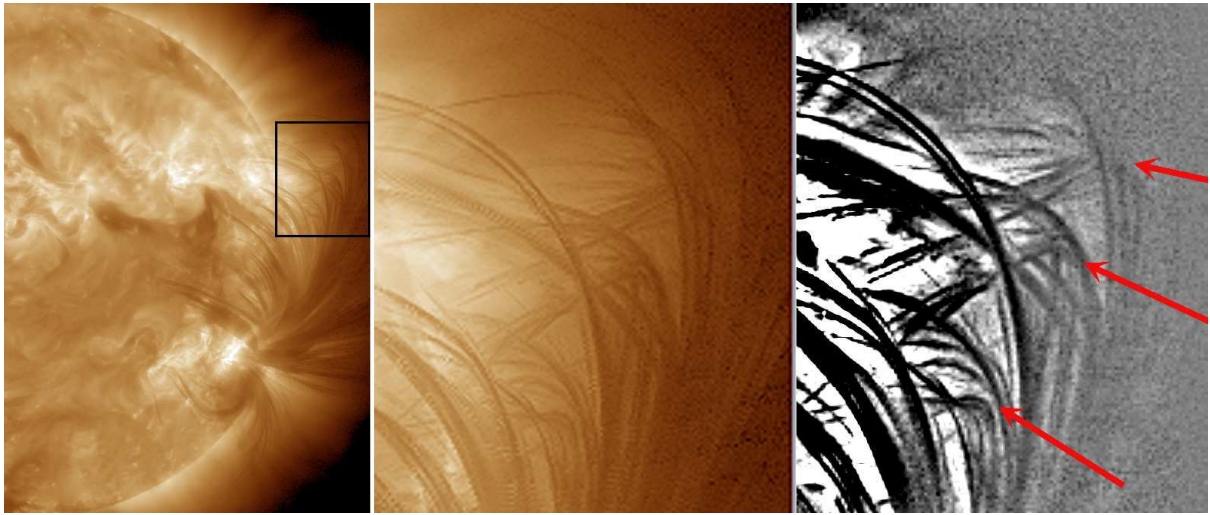}
\caption{\label{fig_images} Persistence-map representation of the trajectories of the moving prominence material during the 2011 June 07 eruption observed by SDO/AIA at 193 {\AA}. The box in the left panel shows the location of the closeup view in the second panel; the third panel is an enhanced version of the second panel. The red arrows point out examples of sharp bends in the trajectories of prominence blobs, which suggest non-ballistic force perturbations. This figure is adapted from \citet{thompson16}.}
\end{center}
\end{figure*}

Figure \ref{fig_images} provides a detailed view of the trajectories of the moving prominence pieces visualized automatically using the persistence-mapping technique \citep{thompson16}. Persistence maps represent a time-integrated history of moving image features, in this case a set of fragmented prominence blobs. The algorithm identified the minimum value of the brightness in each pixel during the lifetime of the eruption event, revealing spatial traces of optically dim features associated with dense prominence plasma \citep{thompson16}. The second and third panels of Figure \ref{fig_images} are zoomed-in versions of the first panel, which provides a global view of the event. 



A close look at the persistence map trajectories provided in Figure \ref{fig_images} reveals multiple occurrences of sharp turns, such as the ones shown with the red arrows, that could indicate detectable non-ballistic perturbations. It is clear that from the persistence map that the features are well-sampled, since their trajectories trace out semi-continuous paths, which offers an opportunity for in-depth analysis of their spatiotemporal behavior and the associated coronal forces.

\section{KODA methodology}
\subsection{Spatiotemporal Feature Tracking}
\label{sec:tracking}

The feature tracking technique used by KODA is based on the spatiotemporal event decomposition method described in previous publications \citep{uritsky10a, uritsky10b, uritsky12, uritsky14, uritsky17}.  The technique is implemented in a three-dimensional space-time defined by the two Cartesian image-plane coordinates and the time axis. The trajectories of the prominence blobs in this three-dimensional domain take the form of lines instead of discrete pieces. Rather than identifying these discrete isolated features in consecutive images, we determined interconnected groups of spatiotemporal pixels that represented each trajectory, and we used these coordinates to recover dynamic characteristics of each plasma blob.

The first step of our methodology consists in the detection of the leading fronts of moving prominence pieces using time-differenced 193 {\AA} AIA images, 
\begin{equation}
\label{eq:dI}
\Delta I(x,y,t) = I(x,y,t-\Delta t) - I(x,y,t),
\end{equation}
where $I$ is the image brightness, $x, y$ are image pixel coordinates, $t$ is the time, and $\Delta t$ is a constant time lag measured in units of the 24-s sampling time. Since the dense and cold prominence material partly blocks the background photon flux coming from the more luminous coronal structures, $\Delta I(x,y,t)$, which represents the difference between the past and the current local luminosity, exhibits a sharp positive spike upon arrival of a leading front of the optically dense prominence material. This material is highly fragmented, and each fragment is characterized by its own propagating front. 

To identify image pixels belonging  the fronts of the moving prominence pieces, the time-differenced images $\Delta I(x,y,t)$ were subject to the adaptive thresholding 
\begin{equation}
\label{eq:th}
\Delta I(x,y,t) > \mu(t) + k \, \sigma_{\Delta I}(t)
\end{equation}
in which $\mu$ and $\sigma_{\Delta I}$ are respectively the mean value and the standard deviation of the differenced image at time $t$, $k$ is a constant dimensionless parameter used to adjust the threshold (the threshold is set at the level of $k$ standard deviations above the mean for each time step). Such adaptive thresholding tends to produce more consistent feature tracking results compared to a fixed detection threshold when the studied data are not sufficiently stationary \citep{klimas17, knizhnik18}.

The image pixels selected by the condition in Eq.\ (\ref{eq:th}) were then assembled into contiguous spatiotemporal clusters using a multidimensional graph-theoretical algorithm presented in earlier publications \citep[see][and references therein]{uritsky10b}. We start by detecting temporal traces of features in individual solar locations, and then verify their spatial adjacency. The method applies hierarchical labeling of clusters of pixels using the ''breadth‐first search'' principle to avoid backtracking of search trees representing individual features. 

\blue{Special precautions have been taken to ensure the quality of tracking results. Small-scale, short-term features which could not correctly represent plasma motions were removed from the analysis (see Section \ref{sec:results} for further details). The trajectories of the remaining features were examined to ensure that they represent systematic long-distance displacements with an expected range of POS velocities. We removed the features whose POS footprint was too large compared to the travelled distance, making the trajectory calculation highly inaccurate. The values of the threshold constant $k$ reported below were chosen to be sufficiently high to make a spatiotemporal overlap of co-moving adjacent plasma blobs unlikely. The tracking algorithm is capable of distinguishing multiple concurrent features with distinct locations or co-spatial features occurring at distinct times \citep[see][for more details]{uritsky13}. However, no tracking method is perfect when applied to real data, and we estimate a small fraction ($< 10 \%$) of the post-selected features to have a more complex morphology due to either a mislabeling of adjacent pixel clusters or an actual merging dynamics of colliding prominence fragments. In future studies, this ambiguity can be further reduced by applying a mass conservation constraint based on time-dependent multi-wavelength column density measurements.
}

Several combinations of detection parameters $(k, \Delta t)$ were tested in order to optimize the performance of the method. Four of these combinations -- (1,2), (1,5), (2,1) and (2,5) -- were found to produce the best feature-tracking results, i.e.\ the maximum number of accurately detected blob fronts with the longest uninterrupted tracking history, at different stages of the eruptive evolution. In our subsequent analysis, we combined the blobs detected using these parameter values into a single statistical ensemble to ensure a reasonably uniform spatial coverage of the eruption region, as discussed below in \S \ref{sec:results}.

Each detected moving feature was represented by a discrete coordinate set $\Lambda_i(t)$ containing time-dependent spatial positions of all image pixels included in the $i^{\text{th}}$ feature at each time step, where $i = 1, ..., N$ is the blob index. After the spatiotemporal domain of the propagating front of each prominence blob was identified based on the $\Lambda_i$ set, the average positions of the moving blobs at different time steps were calculated,
\begin{equation}
\begin{aligned}
\label{eq:avr}
\bar{x}_i(t) &= \frac{1}{N_i(t)}\sum_{j \in \Lambda_i(t)} x_j(t), \\
\bar{y}_i(t) &= \frac{1}{N_i(t)}\sum_{j \in \Lambda_i(t)} y_j(t),
\end{aligned}
\end{equation}
in which $N_i$ is the instantaneous number of spatiotemporal positions occupied by the $i^{\text{th}}$ feature at time $t$, $j$ is the single array index enumerating image pixels, and both coordinates ($x$ and $y$) are measured relative to the Sun's disk center. The statistical uncertainties of the average positions $\bar{x}_i(t)$ and $\bar{y}_i(t)$ were estimated by the corresponding standard deviations, each considered as a function of time,
\begin{equation}
\begin{aligned}
\label{eq:sd}
\sigma_{x,i}^2(t) &= \frac{1}{N_i(t)}\sum_{j \in \Lambda_i(t)} \left( x_j(t) - \bar{x}_i(t) \right)^2, \\
\sigma_{y,i}^2(t) &= \frac{1}{N_i(t)}\sum_{j \in \Lambda_i(t)} \left( y_j(t) - \bar{y}_i(t) \right)^2.
\end{aligned}
\end{equation}
Finally, the time-evolving position vector of each blob was formed, 
\begin{equation}
\vec{r}_i(t) = \left(\bar{x}_i(t), \bar{y}_i(t)\right),
\end{equation}
with the (isotropic) position uncertainty 
\begin{equation}
\sigma_r(t) = \max \left(\sigma_{x,i}(t), \sigma_{y,i}(t)\right)
\end{equation}
defined by the larger of the two standard deviations characterizing uncertainties in the $x$ and $y$ directions.

\subsection{Measuring Trajectories of Prominence Fragments}
\label{sec:dynamics}

The primary goal of the feature-tracking step implemented in KODA is to quantify the contribution of non-central forces to the dynamics of moving prominence blobs. A simplified approach to this problem would be to study the geometry of spatial feature traces to check whether they depart from the elliptical trajectories predicted for purely ballistic motion in the Sun's gravity field. While such a geometric approach could help with determining the presence of non-central forces, it would be unable to measure the magnitude of this force without significant additional assumptions. Adding the time history of the moving features provides an opportunity to evaluate the magnitude of non-central forces from imaging observations under fairly general assumptions, as we show below.

The shape and the time history of the blob tracks can be conveniently linked together by considering perturbations in Kepler's second law, a form of angular momentum conservation commonly used for statistical orbit determinations \citep{tapley04}. This law, which gave the name to our remote-sensing methodology, states that the rate of change of the area $\mathcal{A}$ swept out by the radius vector $\mathbf{r}$ drawn from the center of gravity toward a particle of mass $m$ (the areal velocity), 
\begin{equation}
\frac{d\mathcal{A}}{dt} = \frac{1}{2}r^2\frac{d\theta}{dt},
\label{eq_kepler}
\end{equation}
remains constant if the net torque exerted on the particle is zero, which is the case with any central force. The relevant component of the angular momentum of the particle is given by
\begin{equation}
\mathcal{L}_z =  m r^2 \frac{d\theta}{dt} \bm{\hat{z}}.
\label{eq_ang_mom}
\end{equation}
Comparing Eq.\ (\ref{eq_kepler}) with Eq.\ (\ref{eq_ang_mom}), one gets 
\begin{equation}
\frac{d\mathcal{A}}{dt} = \frac{1}{2m} \vec{\mathcal{L}} \cdot \bm{\hat{z}}, 
\label{eq_cons_mom}
\end{equation}
where $\bm{\hat{z}}=\bm{\hat{r}} \times \bm{\hat{\theta}}$, $\bm{\hat{r}}$ is the instantaneous unit radius vector and $\bm{\hat{\theta}}$ is the unit vector describing the angular coordinate $\theta$. A constant $d\mathcal{A}/dt$ means that the angular momentum is conserved ($\vec{\mathcal{L}} = constant$), which occurs naturally under gravitational forces.

If, however, the areal velocity is not constant, its rate of change reflects a non-zero torque from a non-central force. Differentiating Eq.\ (\ref{eq_cons_mom}) with respect to time and assuming constant mass we obtain
\begin{equation}
\frac{d^2 \mathcal{A}}{dt^2} = \frac{1}{2m} \left(\vec{\mathcal{L}} \cdot \frac{d \bm{\hat{z}}}{dt} + \bm{\hat{z}} \cdot \vec{\tau} \right), 
\label{eq_torque1}
\end{equation}
in which $\vec{\tau} = d \vec{\mathcal{L}} /dt $ is the net torque. 
If the perturbation introduced by $\vec{\tau}$ lies in a single plane, then $\bm{\hat{z}}=constant$ and the torque per unit mass can be obtained directly from the areal acceleration, 
\begin{equation}
\frac{d^2 \mathcal{A}}{dt^2} = \frac{\tau_z}{2 m}.
\label{eq_torque2}
\end{equation}

KODA associates this equation with each fragment of the falling prominence material in order to estimate the individual net torques that are exerted on the fragments. Since the trajectories of the fragments have different orientations relative to the focal plane of the instrument, Eq.\ (\ref{eq_torque2}) should be rewritten in terms of the plane-of-sky (POS) swept-out area $A_i = \mathcal{A}_i \: \text{cos} \: \beta_i$ projected onto the image plane, where $\beta_i$ is the angle between the line of sight (LOS) and the normal to the orbit plane of the $i^{\text{th}}$ fragment. The POS version of Eq.\ (\ref{eq_torque2}) then reads
\begin{equation}
\frac{d^2 A_i}{dt^2} = \frac{\text{cos}\: \beta_i }{2} \frac{\tau_i}{m_i} 
\label{eq_torque3}
\end{equation}

In principle, the orientation angles can be obtained from a three-dimensional tomographic reconstruction of the blob trajectories \citep{thompson09, wood16}. However, such a reconstruction can provide satisfactory results only for a limited subset of prominence blobs reliably observed from two or more platforms. In our present study, we aimed at processing all trackable fragments to estimate torques at the greatest possible number of locations around the erupting prominence. To attain this goal, we approximated the angles $\beta_i$ by constraining the heights of the measured trajectories as described in \S \ref{sec:equations}.

\begin{figure}
\begin{center}
\includegraphics[width=7. cm]{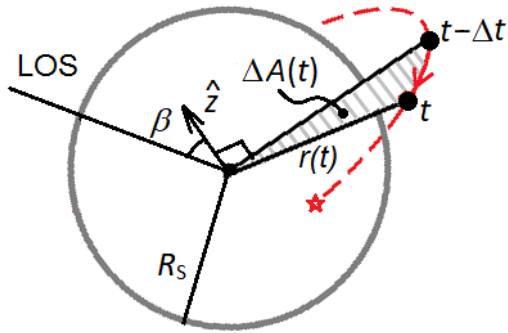}
\caption{\label{fig_sketch} Schematic illustration of the measurement of the increments $\Delta A_i$ of the swept-out area used to evaluate magnetic torques and forces acting on moving pieces of prominence material.}
\end{center}
\end{figure}

To apply Eq.\ (\ref{eq_torque3}) to the SDO/AIA observations of the prominence eruption, we first calculated swept-out area increments $\Delta A_i(t)$ that occurred over each time step (Fig.\ \ref{fig_sketch}). These increments were approximated by the area of the triangle comprised of the projected position $\vec{r}_C = (0,0)$ of the Sun disk center used as the coordinate origin, and the current and previous positions of the moving front of the prominence blob, 
\begin{equation}
\Delta A_i(t) = \frac{1}{2} \left| \vec{r}_i(t) \times \vec{r}_i(t - \Delta t)\right|.
\label{eq_dA}
\end{equation}
The apparent time-dependent area was next approximated by the integral 
\begin{equation}
A_i(t) = \int_{t_{1,i}}^t \Delta A_i(t) dt,
\label{eq_A}
\end{equation}
in which ${t_{1,i}}$ is the time when the $i^{\text{th}}$ blob was first identified in the image sequence. For data filtering purposes, we also computed the tracking lifetime $T_i$ of each feature, 
\begin{equation}
\label{eq_T}
T_i = t_{2,i} - t_{1,i}, 
\end{equation}
where $t_{2,i}$ is the time of the last image containing the blob, and the total linear POS distance, 
\begin{equation}
L_i = \left| \vec{r_i}(t_{2,i}) - \vec{r_i}(t_{1,i}) \right|, 
\label{eq_L}
\end{equation}
 travelled by the detected blob.
 
To measure the acceleration associated with the nonlinear shape of the $A_i(t)$ dependence, we applied fifth order polynomials fits 
\begin{equation}
A_i(t) \approx a_i + b_it + c_it^2 + d_it^3 + e_it^4 + f_it^5. 
\label{eq_poly}
\end{equation}
The polynomial coefficients $a_i$, $b_i$, $c_i$, $d_i$, $e_i$ and $f_i$ were computed using a least-mean-square optimization algorithm based on matrix inversion implemented by the {\it POLY\_FIT} function of the {\it Interactive Data Language (IDL)}. The fifth order of the polynomial approximation enabled
analysis of the nonlinear time evolution of the areal acceleration, expressed by the cubic function 
\begin{equation}
\frac{d^2 A_i}{dt^2} \approx 2c_i + 6d_it + 12e_it^2 + 20f_it^3. 
\label{eq_areal_acc}
\end{equation}
This nonlinear form allows for changes of sign of the torque along the blob trajectory, as discussed later in the text. Figure \ref{fig_examples} presents several examples of the time evolution of the blob area $A_i$ and its acceleration; the red lines show the polynomial fits for each of these blobs.

\subsection{Calculating Physical Parameters}
\label{sec:equations}

The measured areal accelerations were used to evaluate the non-central force densities exerted on each of the fragments of prominence material tracked by our code. 
In the low-beta environment of the corona, a non-zero right hand side of Eq.\ (\ref{eq_torque3}) could result from the transverse component of the 
Lorentz force exerting a torque on the rising and falling prominence material. Since both the mass and the viewing cosine angle are always positive, a negative (positive) areal acceleration implies a decreasing (increasing) total angular momentum, which can be interpreted respectively as magnetic braking and magnetic acceleration.  Oscillations of the areal acceleration along the trajectory of a falling prominence fragment could indicate the presence of MHD waves and/or periodic magnetic structures.

In our calculations, we used the column (area) mass density $\rho_{col}\approx 4.3 \times 10^{-4}$ kg m$^{-2}$ reported by \citet{gilbert13} to calculate volumetric mass densities
\begin{equation}
\rho_i = \rho_{col} \left\langle S_i(t)^{-1/2} \right\rangle_t,    
\label{eq_rho}
\end{equation}
where $S_i(t)$ is the time-evolving POS area of the $i^{\text{th}}$ prominence blob and $\left\langle... \right\rangle_t$ denotes averaging over all the time steps $t \in [t_{1,i}, t_{2,i}]$ at which the blob was observed. The sizes and volumes of the blobs were evaluated assuming, respectively, 
\begin{equation}
\begin{aligned}
D_i & = \left\langle S_i(t)^{1/2} \right\rangle_t, \\
V_i & = D_i^3, \\
\end{aligned}
\end{equation}
and the masses were obtained from
\begin{equation}
m_i = \rho_{col} \left\langle S_i(t) \right\rangle_t.
\end{equation}

The angles $\beta_i$ between the LOS and the normal to the orbit plane were approximated by the ratio of the projected POS area of the triangle comprised of the center of the Sun and the first and the last positions of the blob to the true area of the same triangle, 
\begin{equation}
\cos \beta_i = \left| \frac{ \vec{r}_i(t_{1,i}) \times \vec{r}_i(t_{2,i})}{ \vec{r}\,'_i(t_{1,i}) \times \vec{r}\,'_i(t_{2,i})}\right|, 
\end{equation}
in which 
\begin{equation}
\vec{r}\,'_i = \left(x_i(t), y_i(t), \sqrt{r^2 - r_i^2}\right) 
\end{equation}
is the reconstructed three-dimensional position vector whose $z$ component was obtained under the assumption that the tracked blob trajectories are constrained to a nominal spherical surface concentric to the Sun and described by the radius 
\begin{equation}
r = R_S + h, 
\end{equation}
where $h \approx 0.2\, R_S$ is the average estimated altitude of tracked prominence blobs and $R_S \approx 7 \times 10^8\,$m is the solar radius. 

The angular accelerations of the blobs were computed based on the apparent areal accelerations (Eq.\ \ref{eq_torque3}) and the moments of inertia $m_i r^2$ describing each blob treated as a point mass, \begin{equation}
\alpha_i = \frac{2}{r^2\cos \beta_i}\frac{d^2A_i}{dt^2}.
\label{eq_ang_acc}
\end{equation}
The magnitudes of the magnetic torques acting on the prominence debris were obtained from
\begin{equation}
\tau_{B, i} = m_i r^2 |\alpha_i|,
\label{eq_B_torque}
\end{equation}
and the magnetic force densities associated with these torques were evaluated using
\begin{equation}
f_{B, i} = \frac{\tau_{B,i}/r}{V_i} \equiv \rho_i \, r \, |\alpha_i|.  
\label{eq_B_force}
\end{equation}

The obtained magnetic force estimates were next used to estimate the strength of the magnetic field passing through each plasma blob. In this calculation, we assumed $f_B$ to be the magnetic tension force created by curved magnetic field lines guiding the motion of the frozen-in plasma material, with the average curvature radius $R_{c,i}$ estimated for the trajectory $\vec{r}\,'_i(t)$ of the $i^{\text{th}}$ blob, 
\begin{equation}
B_i \approx (\mu_0 R_{c,i} f_{B, i})^{1/2}.
\label{eq_B_field}
\end{equation}
\blue{The curvature was calculated by approximating the reconstructed three-dimensional trajectories $\vec{r}\,'_i$ of the blobs by circular arcs.} Using the obtained values of $B_i$ and the measured speeds $v_i$ of the blobs, we computed the magnetic ($p_{B, i}$), dynamic ($p_{dyn, i}$), and thermal ($p_{th, i}$) pressures of the descending prominence plasma, 
\begin{equation}
\begin{aligned}
p_{B,i} &= \frac{B_i^2}{2 \mu_0}, \\ 
p_{dyn,i} &= \frac{1}{2} \rho_i v_i^2, \\ 
p_{th, i} &= 2 \frac{\rho_i}{m_p} k_B T_b. \\
\end{aligned}
\end{equation}
To estimate the thermal pressure, we used the ideal-gas law in which $m_p$ is the proton mass, $k_B$ is Boltzmann's constant, and $T_b$ = 3.3$\times$10$^4$ K is the blob temperature reported for the event by \citet{landi13}. 
In addition to these pressure parameters, the volumetric density $u_{g}$ of the gravitational potential energy relative to the solar surface was approximated by the value 
\begin{equation}
u_{g,i} \approx \rho_i g_S h, 
\label{eq_grav_energy}
\end{equation}
where $g_S$ = 2.75$\times$10$^2$ m s$^{-2}$ is the gravitational acceleration at the Sun's surface.

\section{Testing and Results}
\label{sec:results}

The spatiotemporal feature tracking algorithm presented in \S \ref{sec:tracking} enabled the detection of numerous individual prominence fragments during the interval 06:30 - 8:00 following the prominence eruption. By varying the floating threshold coefficient $k$ (Eq.\ \ref{eq:th}) as well as the time lag $\Delta t$ used for constructing time-differenced images (Eq.\ \ref{eq:dI}), we were able to track on the order of $10^2$ prominence pieces for each combination of parameters. The blobs were next filtered by requiring that two conditions, 
\begin{equation}
T_i \geqslant 240 \,\, \text{s}, \,\,\,\,  L_i \geqslant 35 \,\, \text{Mm},    
\label{eq_filt}
\end{equation}
 are simultaneously fulfilled. The first of these conditions eliminated short-lived features that likely resulted from a detection error; the second removed quasi-static features not associated with the erupting prominence. The statistics of the tracking results are summarized in Table \ref{tab:tracking}, in which $n_{tot}$ and $n_{filt}$ are respectively the numbers of prominence blobs before and after the filtering condition (Eq.\ \ref{eq_filt}) is applied, and $\overline{T}$ and $\overline{L}$ are the ensemble-averaged $T_i$ and $L_i$ values calculated for the filtered populations of events.

\begin{deluxetable}{cc|cccc}
\tablecaption{Basic statistics of the tracked prominence blobs\label{tab:tracking}}
\tablewidth{0pt}
\tablehead{
\colhead{$\Delta t$} & \colhead{$k$} & \colhead{$n_{tot}$} & \colhead{$n_{filt}$} & \colhead{$\overline{T}$, s} & \colhead{$\overline{L}$, Mm}  
}
\decimalcolnumbers
\startdata
   2 & 1 & 104 & 29 & 586 & 57 \\
    2 & 2 & 89 & 50 & 846 & 88 \\
    5 & 1 & 111 & 40 & 692 & 73 \\
    5 & 2 & 86 & 52 & 872 & 76 \\
\enddata
\end{deluxetable}

The different choices of $\Delta t$ and $k$ reflected in  Table \ref{tab:tracking} focused the tracking algorithm on substantially different populations of prominence pieces. Varying $k$ by a factor of 2 tuned the algorithm to select prominence blobs characterized by significantly different absorption levels relative to the background coronal emission. As Table \ref{tab:tracking} shows, a larger $k$ (higher absorption) reduces the total number of detected events while increasing the fraction of tracking results meeting the filtering conditions (Eq.\ \ref{eq_filt}). In turn, changing the time lag leads to an identification of blobs characterized by substantially different leading front steepness, with the $n_{filt}/n_{tot}$ ratio tending to be larger for $\Delta t = 5$ allowing for more gradual blob fronts, compared to $\Delta t = 2$ requiring sharper fronts. In the subsequent analysis, we combined the features obtained with $k = 1, 2$ and $\Delta t = 2, 5$ into a single statistical set in order to maximize the spatial coverage of the coronal volume affected by the prominence eruption.

\begin{figure*}

\hspace{1.5in} (a) \hspace{2.5in} (b)

\begin{center}
\includegraphics[height=6 cm]{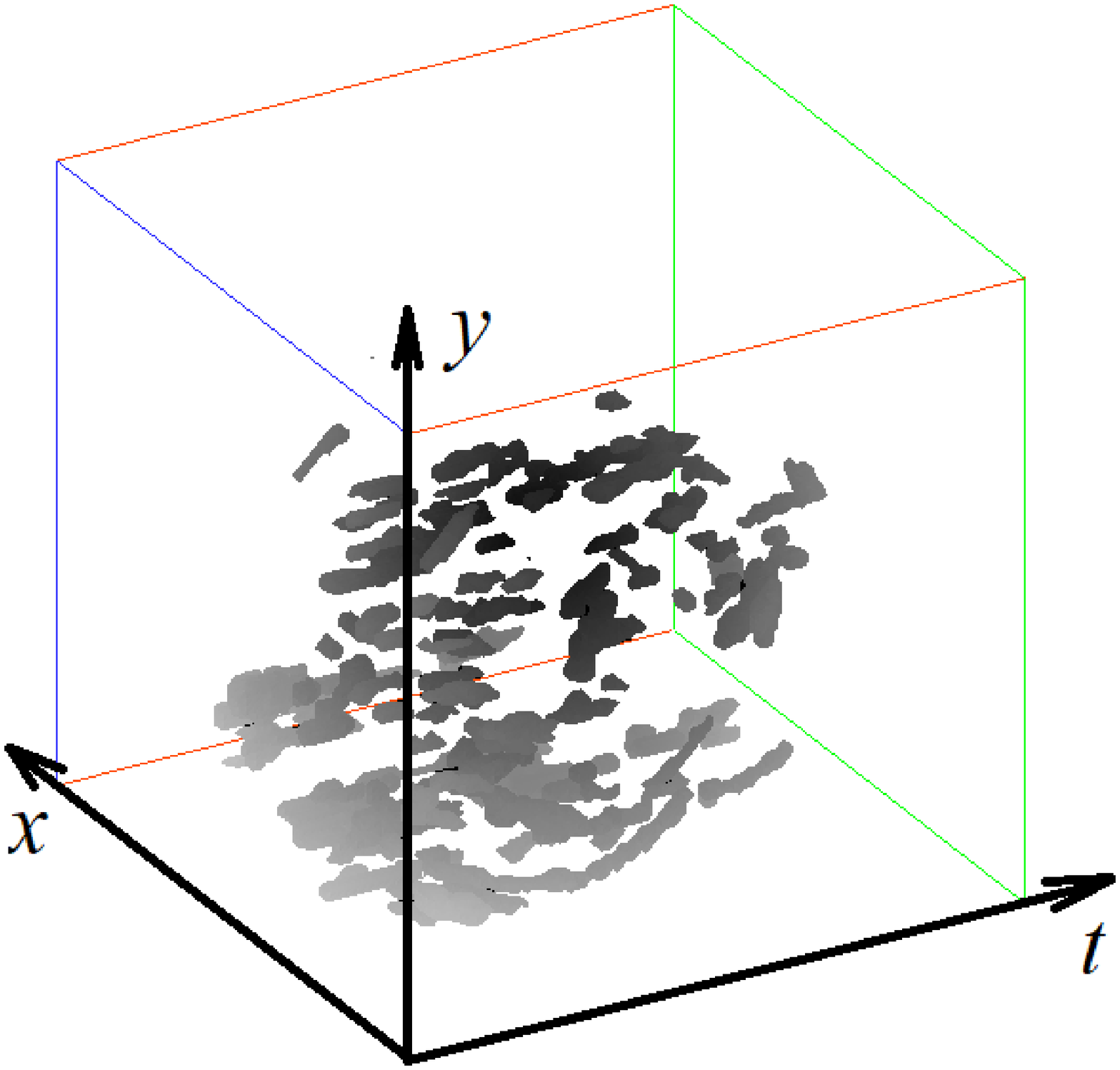}
\includegraphics[height=6 cm]{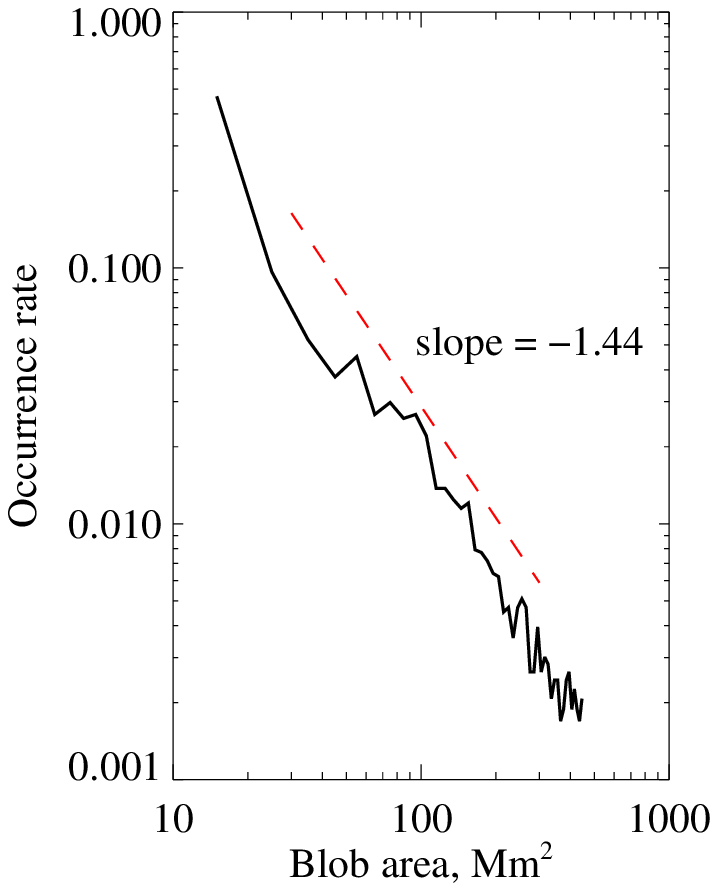}
\end{center}
\hspace{1.5in} (c) \hspace{1.3in} (d) \hspace{1.3in} (e)  
\begin{center}
\includegraphics[height=5.5 cm]{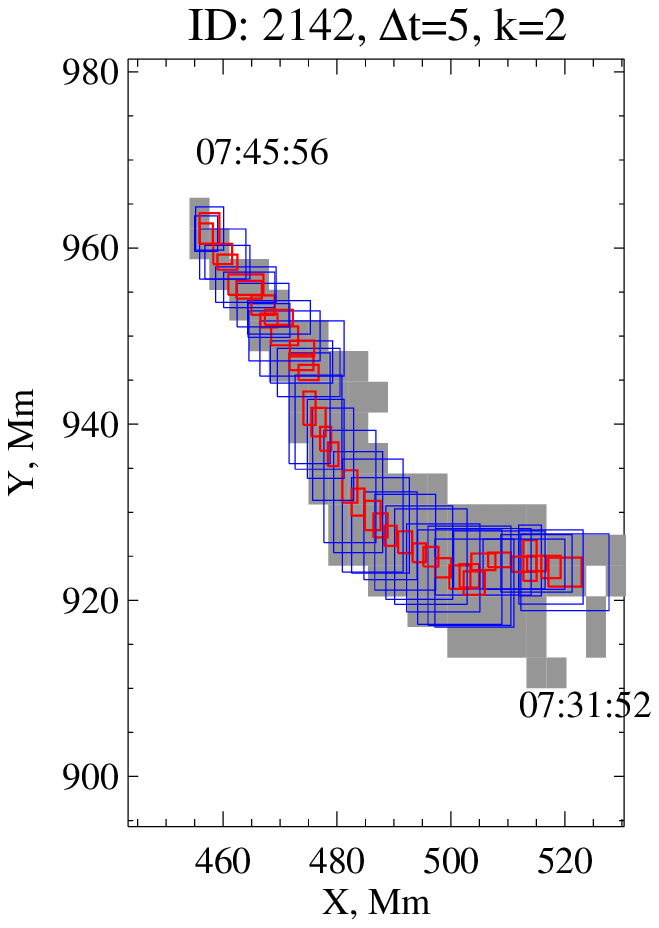}
\includegraphics[height=5.4 cm]{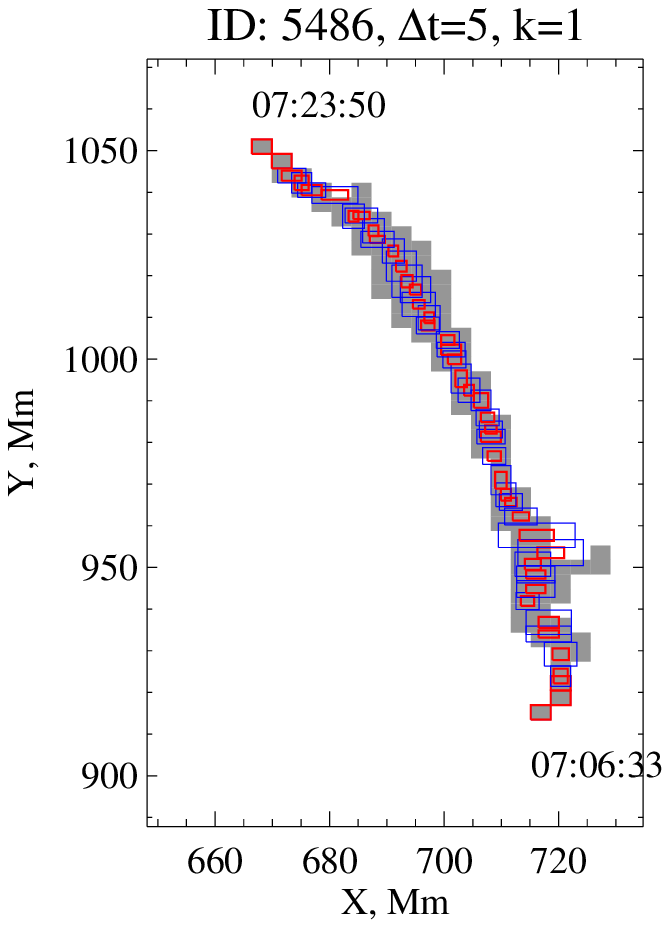}
\includegraphics[height=5.5 cm]{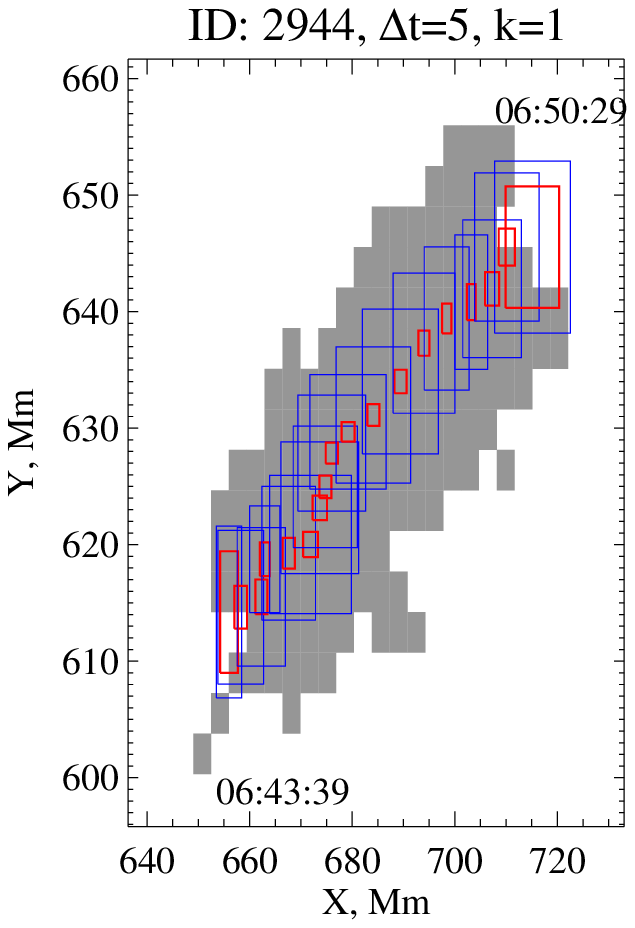}

\caption{\blue{Illustration of the KODA feature-tracking methodology. (a) The spatiotemporal volume in which the tracking is performed. Gray shapes are the time-dependent pixel clusters $\Lambda_i$ detected using $\Delta t = 5$ and $k=1$, see Section \ref{sec:tracking} for details. Smallest clusters with spatiotemporal volumes less than 30 pixels are removed for clarity. (b) Probability distribution of instantaneous blob areas for all detected events, for all blob indices $i$ and time steps $t$. (c-e) Characteristic examples of prominence blob trajectories tracked by the code. Shaded grey area is the footprint of each track in the AIA image plane. The blue rectangles show the time-dependent standard deviations (Eq.\ \ref{eq:sd}) describing statistical uncertainties of the average positions $\bar{x}_i(t)$ and $\bar{y}_i(t)$ (Eq.\ \ref{eq:avr}) of the moving blobs; red rectangles are the corresponding standard errors. Initial ($t_{1,i}$) and final ($t_{2,i}$) tracking times defining the blob lifetime $T_i$ (Eq. \ref{eq_T}) are provided for each example. \label{fig_feature_tracking} } }.
\end{center}
\end{figure*}

\blue{
Figure \ref{fig_feature_tracking} explains the methodology used for the detection of the moving plasma blobs in the studied stack of AIA images. Figure \ref{fig_feature_tracking} (a) shows the locations and shapes of spatiotemporal clusters identified using one of the four combinations of tracking parameters listed in Table \ref{tab:tracking}. Previous applications of the tracking algorithm to multidimensional data sets have demonstrated its ability to unambiguously separate clusters embedded in a volume even when the features overlap along individual coordinate axes \citep{uritsky10b}. The probability histogram of blob areas shown on Figure \ref{fig_feature_tracking}(b) exhibits an approximate power-law form, with a negative log-log slope of about 1.4. The median value of the population is about 24 Mm$^2$, suggesting that a typical linear size of the tracked blob is around 4-5 Mm. Heavy-tailed distributions such as the one shown here are indicative of a branching stochastic process in which an initially ordered system undergoes a cascade of splitting events leading to a growing number of ever smaller fragments. Such cascade could be a signature the Rayleigh-Taylor instability \citep{carlyle14, innes12}, although the relatively high values of the magnetic field strength (tens of Gauss) reported below could prevent this scenario. Panels (c)-(d) of the same Figure provide characteristic examples of spatial traces of moving prominence blobs tracked by our code, as well as the statistical distribution describing the occurrence rate of blob areas $S_i(t)$ for all studied blobs. The solid grey polygons in the blob examples show the image regions visited by each blob during its lifetime $T_i$. The blue and red rectangles show, respectively, the standard deviation and standard error in the determination of the blob positions at each time step. The standard errors characterizing coordinate uncertainties are typically much smaller than the non-ballistic perturbations to the blob trajectories caused by the coronal magnetic field and targeted by KODA. 
}

\begin{figure*}
\begin{center}
\includegraphics[height=7.5 cm]{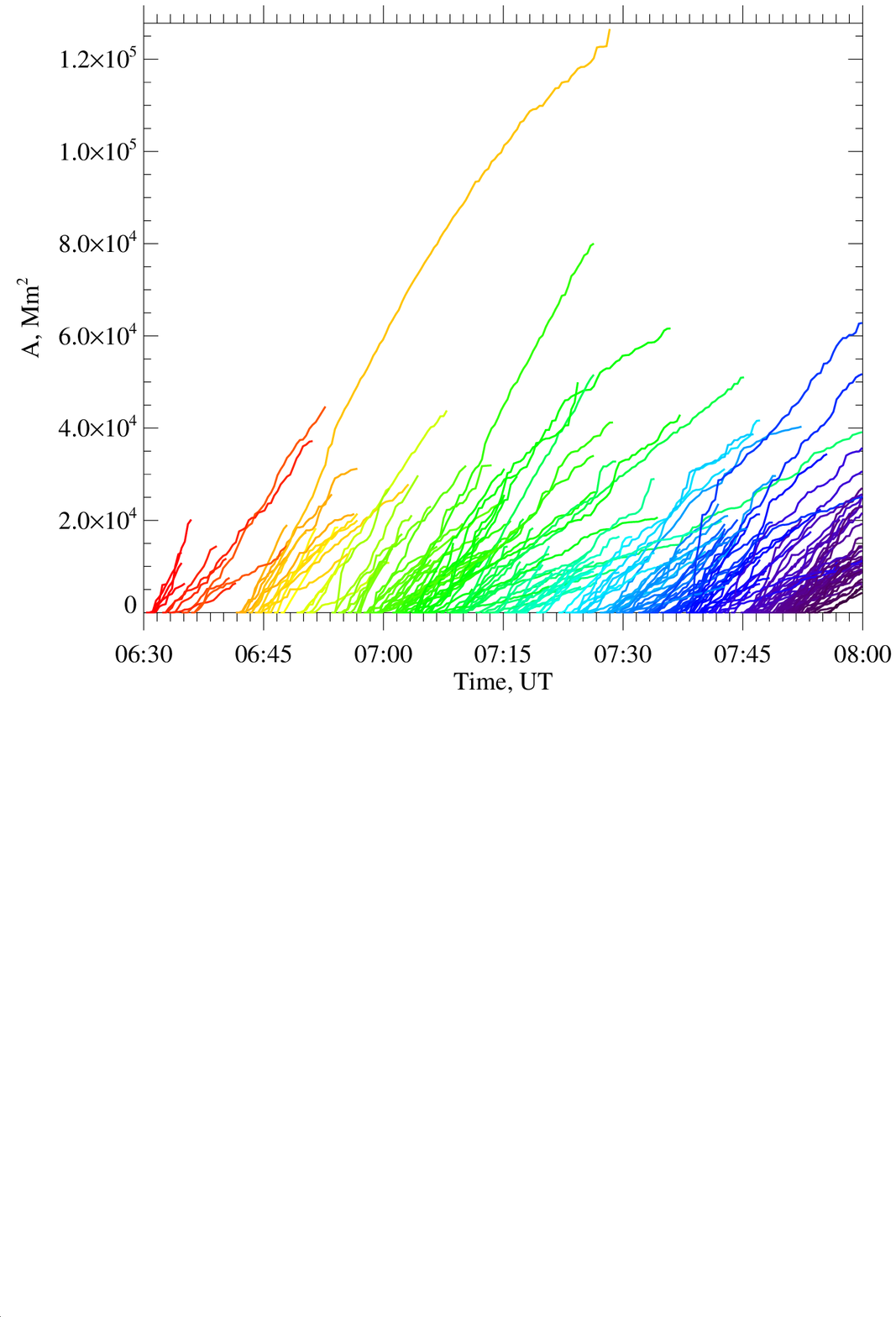}\includegraphics[height=7.5 cm]{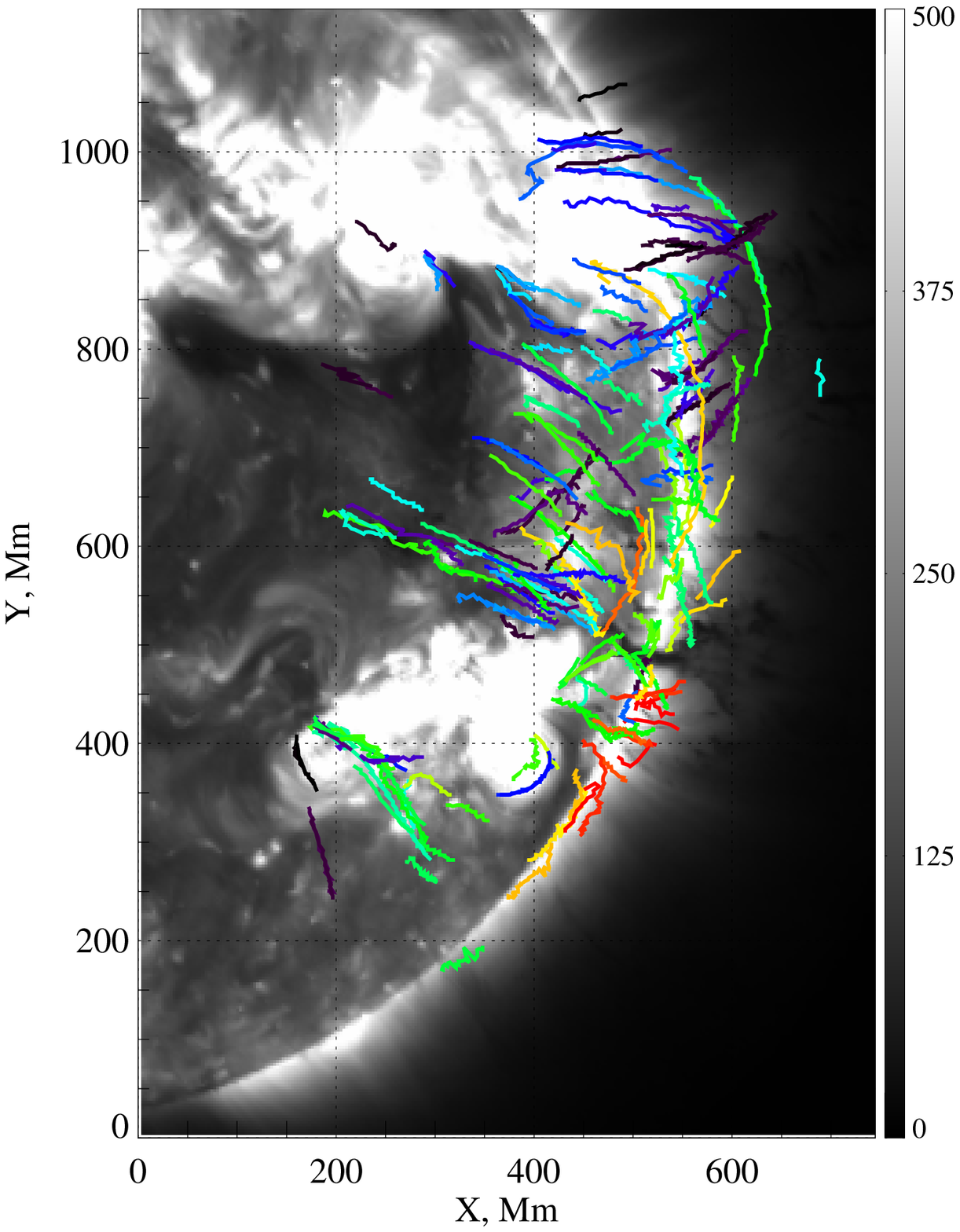}
\caption{Left: Time evolution of the estimated swept-out areas $A_i$ of plasma blobs tracked during the 2011 June 07 eruption event. A purely ballistic motion results in a linear $A_i(t)$ dependence. It is evident that many of the blobs do not obey this rule. Right: spatial trajectories of the detected blobs overplotted with an AIA 193 {\AA} image taken from the middle of the studied time interval. In both panels, the color encodes the starting time of the feature on a scale ranging from red (6:30 UT) to violet (8:00 UT). \label{fig_events} }.
\end{center}
\end{figure*}

\begin{figure*}
\begin{center}
\includegraphics[width=15.0 cm]{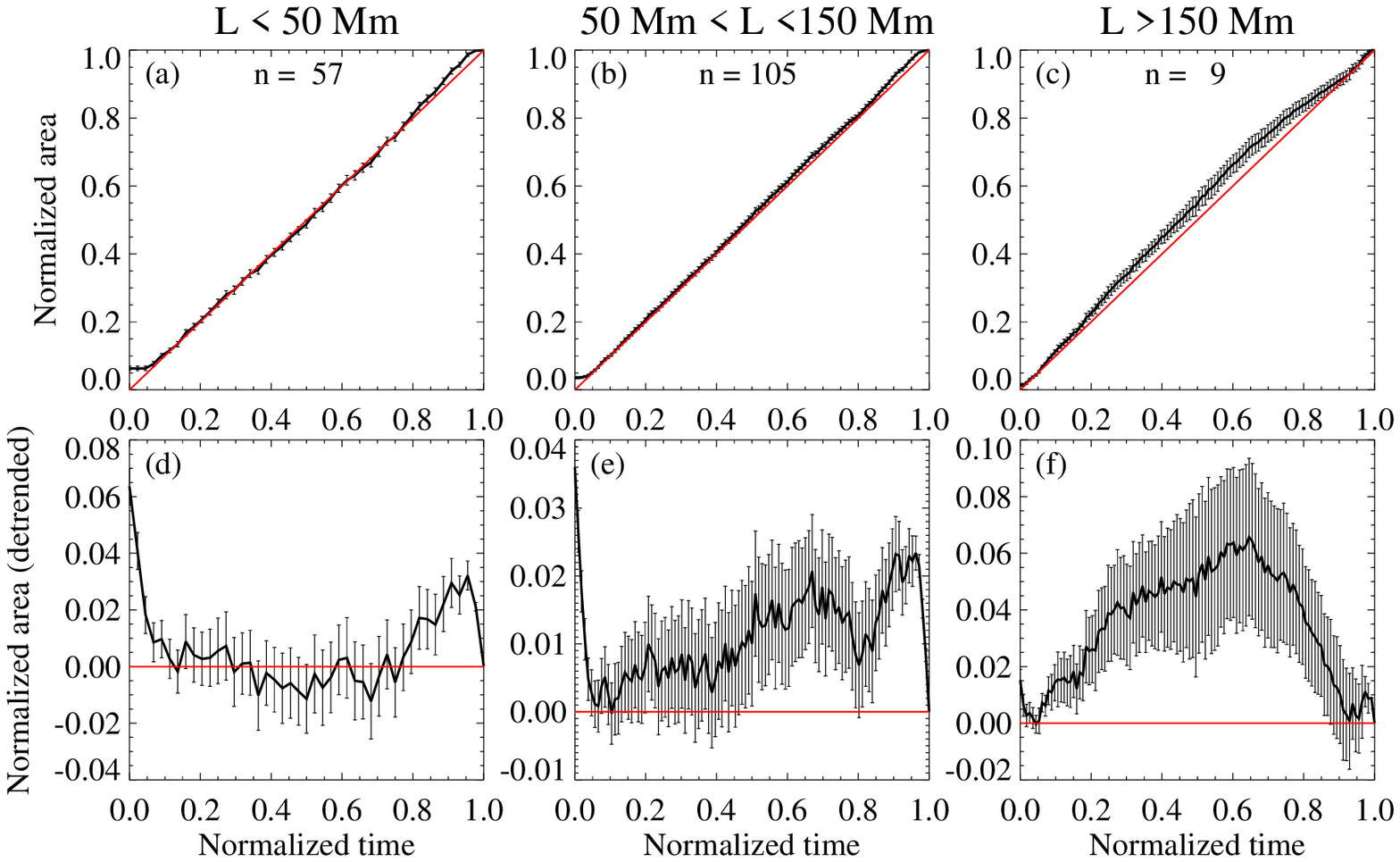}
\caption{\label{fig_superposed} Superposed epoch analysis of the temporal signatures of the swept-out areas characterizing detected prominence blobs. Top row (a-c): ensemble-averaged normalized swept-out area versus normalized tracking time for three different ranges of the POS distance $L_i$ travelled by blobs. The number of plasma blobs in each $L$ interval is shown on each panel. Solid red lines indicate linear areal evolution with zero angular acceleration. Bottom row (d-f): detrended versions of the plots from the top row obtained by subtracting the linear trends. Vertical bars are the standard errors of each superposed epoch average.}
\end{center}
\end{figure*}

Figure \ref{fig_events} shows the temporal evolution of the apparent swept-out areas $A_i$  of the detected coronal features (left panel) and their spatial trajectories as described by $\vec{r}_i(t)$ (right panel). The line color on both panels communicates information about the starting time $t_{1,i}$ of the feature on the scale from red (6:30 UT) to dark blue (8:00 UT). The same color coding is used in Figures 6-9 discussed later in the text.

It can be seen that many of the blobs in Figure \ref{fig_events}, including the blob with the longest tracked history initiated around 6:45 UT, exhibit a nonlinear growth of $A(t)$ signaling the effect of a non-zero torque. This torque also influences the spatial tracks of the features by perturbing the elliptical shapes expected for purely ballistic motion. Theoretically, the non-ballistic perturbations should appear in both spatial and temporal analysis domains, but we found that the areal dynamics is more reliable as an empirical non-ballistic indicator compared to the spatial or temporal signatures alone. 

To test the statistical significance of the nonlinearity of the swept-out area evolution measured by KODA, we performed a superposed epoch analysis of all detected prominence blobs. The duration and the dynamic range of each $A_i(t)$ curve were first rescaled to fit the interval $[0,1]$, after which ensemble-averaged rescaled curves were computed for several ranges of the travelled distance $L_i$ (Eq.\ \ref{eq_L}) approximating the length of the track. The upper panels, Figure \ref{fig_superposed} (a-c), show the results of this averaging for three different ranges of $L_i$. The curved black line on each panel is the ensemble-averaged rescaled swept-out areas plotted versus the rescaled time, the vertical bars are the standard errors of the plotted mean values, and the red diagonal line connecting the points $(0,0)$ and $(1,1)$ corresponds to the linear growth of the swept-out area. The bottom panels, Figure \ref{fig_superposed} (d-f), present the differences between the ensemble-averaged rescaled areas and the linear model in each $L$ range. It can be seen that the difference between the two lines often exceeds the standard error, signaling that the non-ballistic perturbations are statistically significant. 

\begin{figure*}
\begin{center}
\includegraphics[width=6.0 cm]{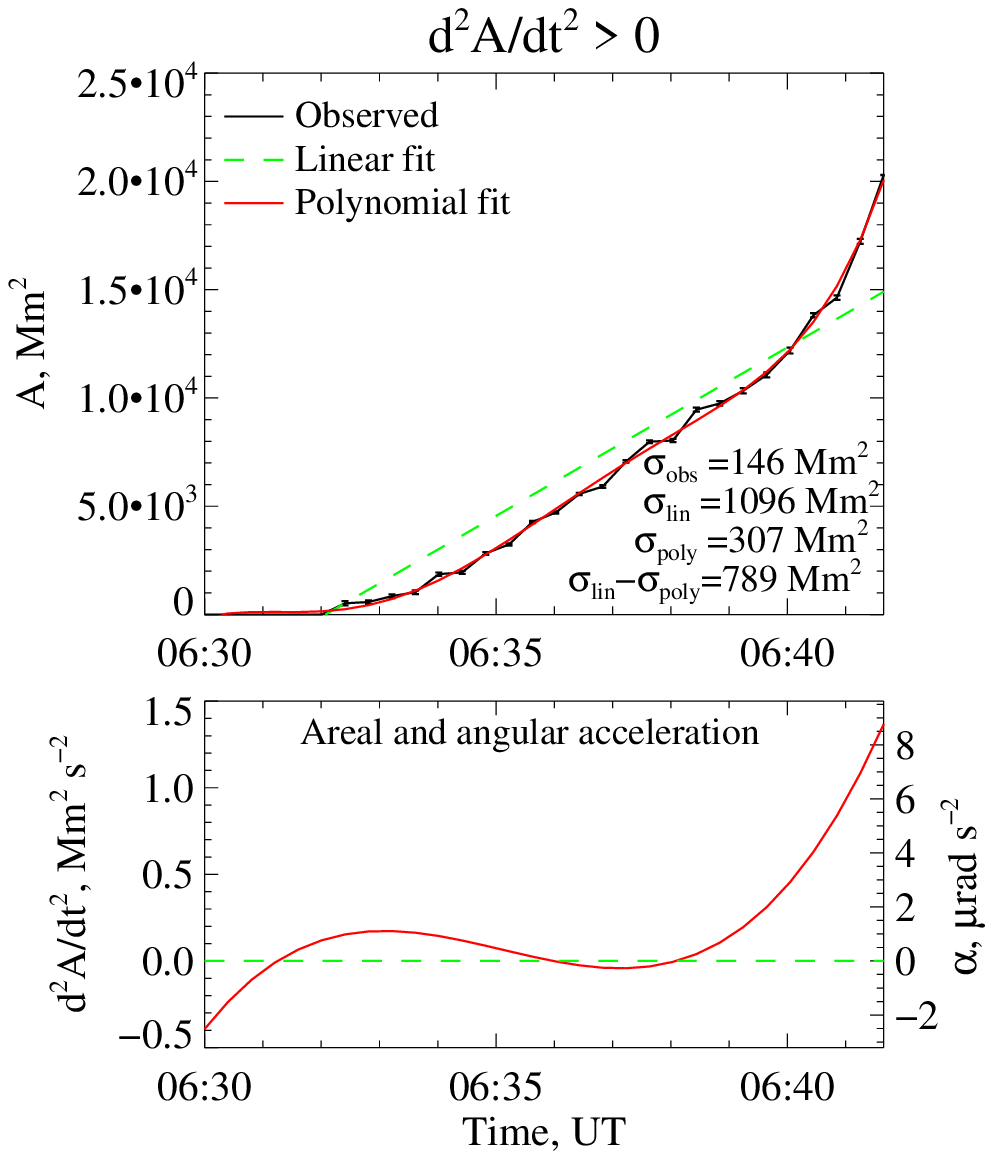}\includegraphics[width=6.0 cm]{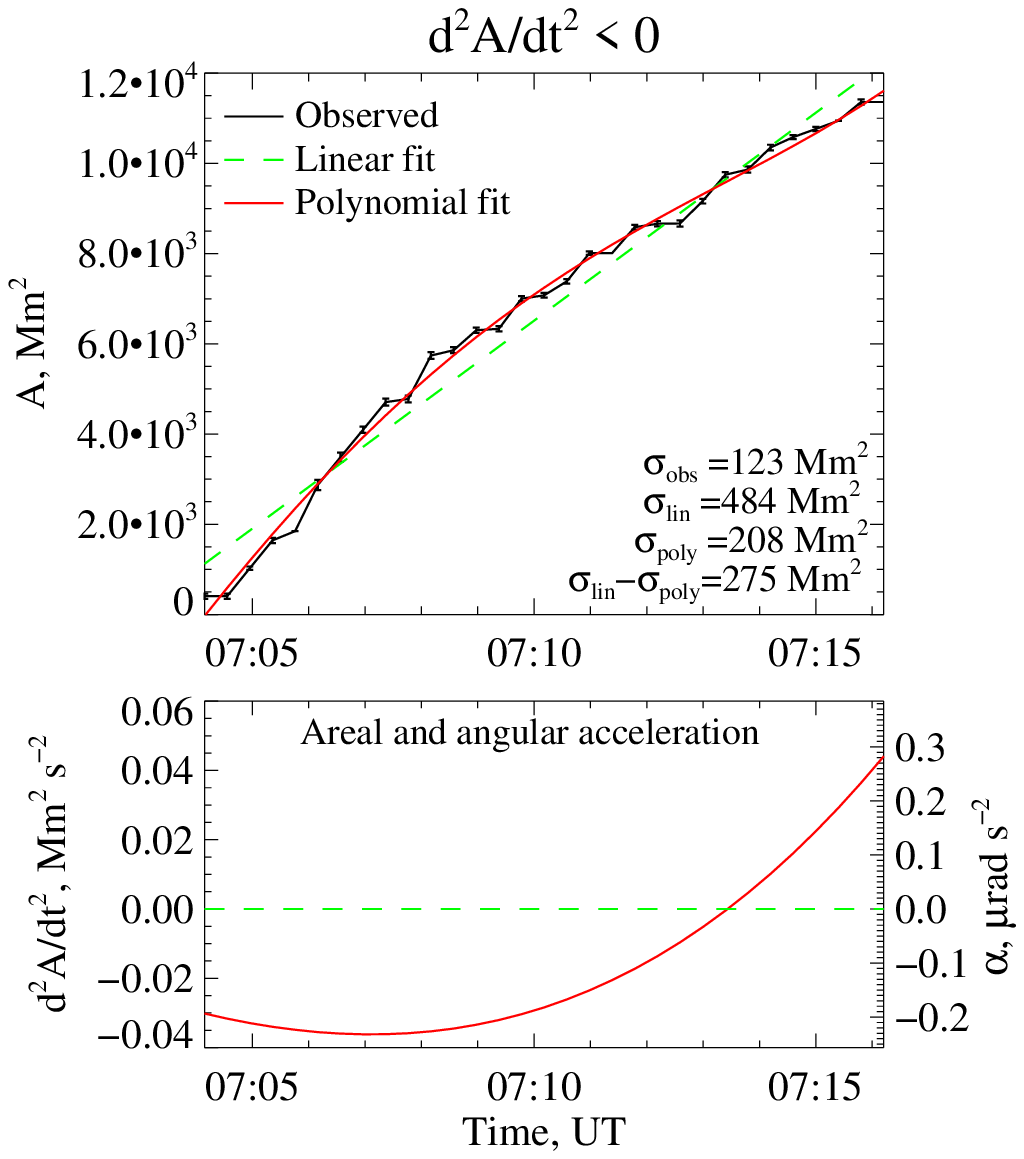}\includegraphics[width=6.0 cm]{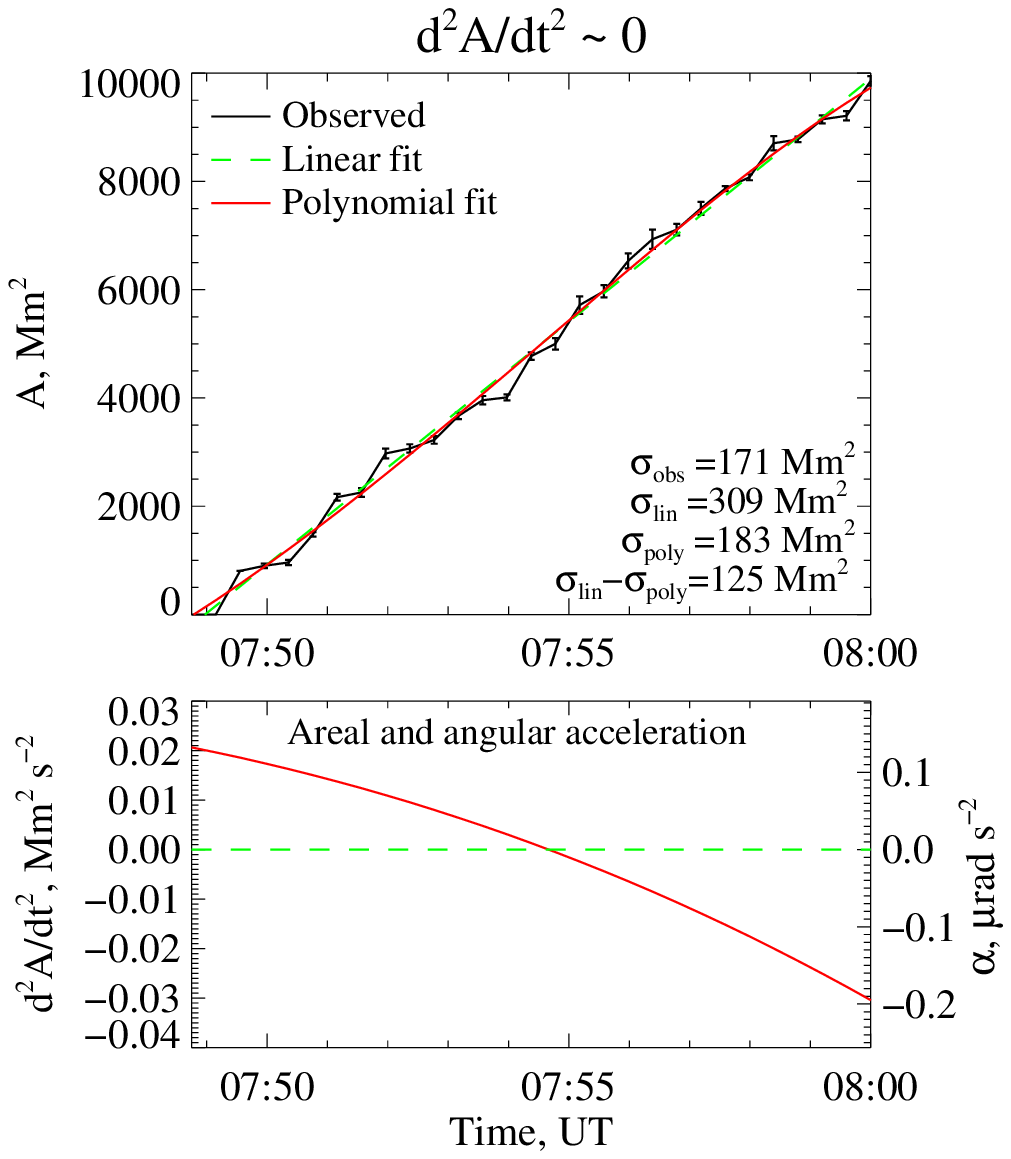}
\caption{Examples of prominence blobs exhibiting positive (left), negative (center), and near-zero (right) net accelerations. Top row: black curves show measured swept-out areas as a function of time in event; dashed green and solid red lines are respectively linear and polynomial fits to the measured areas. The provided uncertainty values are used to identify cases with statistically significant non-linearity, as described in the text. Bottom panels: areal acceleration (left axes) and the corresponding angular acceleration (right axes) computed using the polynomial fit. \label{fig_examples} }.
\end{center}
\end{figure*}

The concave shape of the superposed epoch evolution of plasma blobs with $L_i > 50$ Mm (Figure \ref{fig_superposed}(f)) could indicate the presence of a large-scale magnetic braking affecting the dynamics of longest prominence tracks in our data. At shorter scales, however, the coronal magnetic field can lead to both braking and acceleration signatures. Figure \ref{fig_examples} provides three examples of swept-out area signatures of prominence blobs characterized by positive, negative, and near-zero acceleration. The upper row of panels shows the time evolution $A(t)$, which demonstrates respectively a convex (left), concave (center), and quasi-linear (right) shape. The measured swept-out areas (black lines) are overplotted with 5$^{\text{th}}$ order polynomial fits (red lines, Eq.\ \ref{eq_poly}), as well as with linear least-square fits (dashed green lines) provided for comparison. 

The statistical significance of the nonlinear shape of $A(t)$ was tested by comparing the propagated observational error $\sigma_{obs}$ describing the swept-out area measurements with the difference between the fitting errors $\sigma_{poly}$ and $\sigma_{lin}$ obtained from the polynomial and linear models, correspondingly. The first and second examples (left and center panels, Fig.\ \ref{fig_examples}) are characterized by the condition $\sigma_{lin} -\sigma_{poly} > \sigma_{obs}$. This ensures that the gain in the fitting accuracy achieved by nonlinear fitting exceeds the observational error, and that the nonlinearity of the $A(t)$ shape is significant. The third example (right panels, Fig.\ \ref{fig_examples}) is described by $\sigma_{lin} -\sigma_{poly} < \sigma_{obs}$, which indicates that the swept-out area evolution in that case was linear within the accuracy of our measurements.
 
The lower row of panels in Figure \ref{fig_examples} presents the evolution of the second time derivative of the swept-out area (Eq.\ \ref{eq_areal_acc}) obtained from the polynomial fits. The right axes are calibrated in terms of the angular acceleration computed using Eq.\ (\ref{eq_ang_acc}). In the first example (Fig.\ \ref{fig_examples}, left), the areal acceleration $d^2 A/dt^2$ remains positive during most of the tracking lifetime $T_i$, which implies a positive out-of-plane torque component $\tau_z$ as defined by Eqs.\ (\ref{eq_torque2}) and (\ref{eq_torque3}). Since this prominence fragment was observed at the early stage of the eruption process, it is reasonable to assume that the detected torque was caused by the magnetic tension force exerted by the recently reconnected loop system, which likely possessed a significant initial field line curvature.

The second example (Fig.\ \ref{fig_examples}, center) exhibits a consistently negative acceleration ($d^2 A/dt^2<0$), except for the last couple of minutes of the tracked evolution. The magnitude of the areal acceleration experienced by this plasma blob is substantially smaller than that detected in the first example, indicating a weaker magnetic force contributing to the non-ballistic evolution of the second event. 
The last example (Figure \ref{fig_examples}, right) demonstrates moderate values of positive (negative) acceleration during the first (second) half of lifetime of this blob event. Within the measurement uncertainty, the time-averaged acceleration of this plasma blob was zero, in agreement with our earlier conclusion about the linear form of its areal dynamics resulting from the error analysis. 

\begin{figure*}
\begin{center}
\includegraphics[width=17.0 cm]{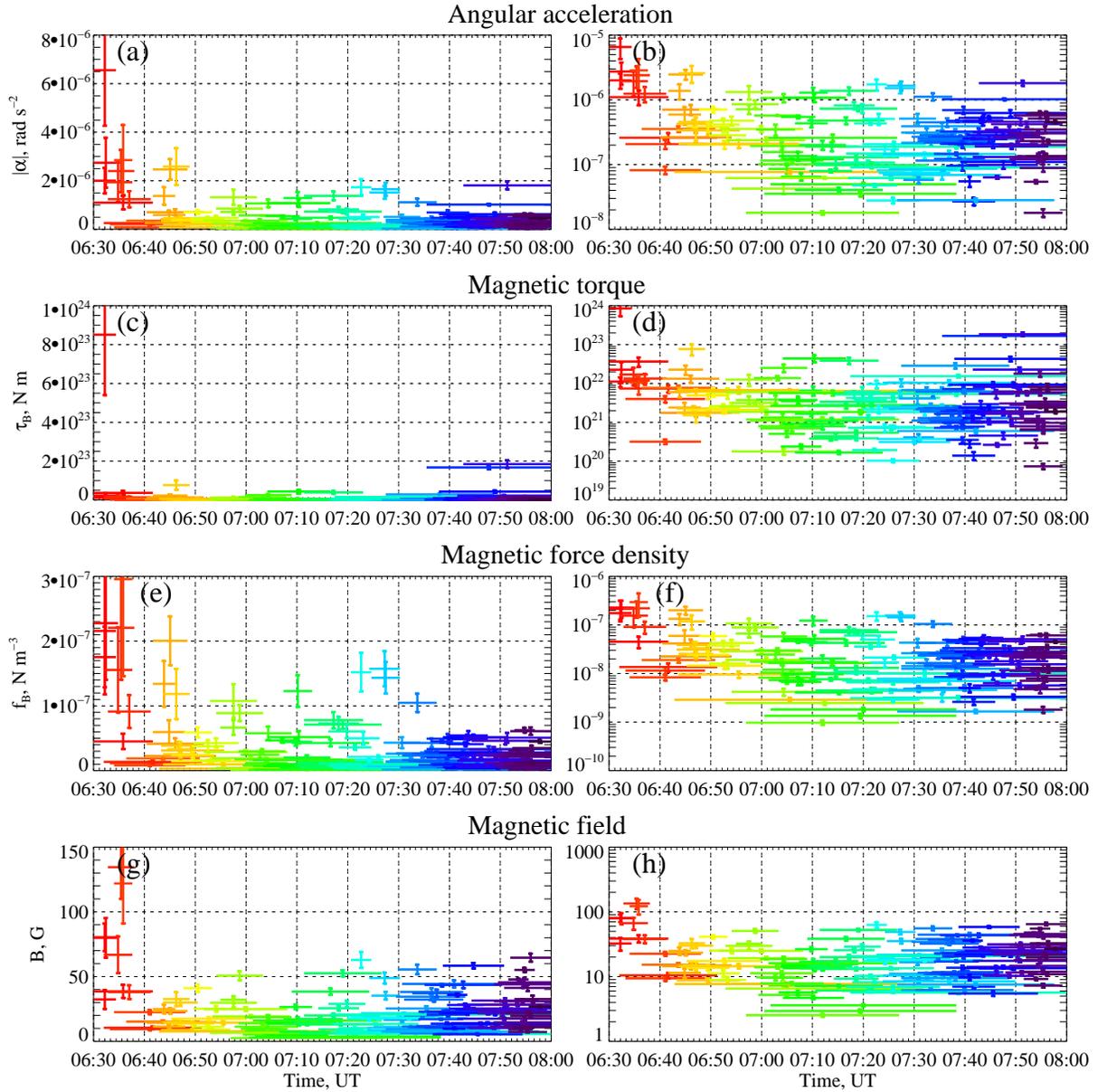}
\caption{\label{fig_dynamics} Physical characteristics of falling prominence material estimated by KODA  during the 2011 June 07 event. From top to bottom: angular acceleration of prominence blobs estimated based on the time evolution of the swept-out areas, magnetic torque associated with the measured angular acceleration, and the magnetic tension force and the magnetic field strength consistent with the estimated torque, plotted on linear (left) and logarithmic (right) scales. Horizontal and vertical bars show the time interval of each tracking event and the standard errors of each measurement, respectively. Color coding is the same as in Figure \ref{fig_events}. }
\end{center}
\end{figure*}

Figure \ref{fig_dynamics} shows a set of average physical characteristics of prominence blobs, estimated using Eqs.\ (\ref{eq_ang_acc}-\ref{eq_B_field}), based on the observations of the swept-out area evolution of each blob. The left and right columns of panels use respectively linear and logarithmic vertical axes. The horizontal bars on each plot mark the time intervals $[t_{1,i}; t_{2,i}]$ over which the blobs were continuously tracked, and the vertical bars are the standard errors of the measurements.  The color coding reflects the sequential order of the features, as determined by their start times $t_{1,i}$, and is the same as that used in Figure \ref{fig_events}.

Figure \ref{fig_dynamics}(a-b) shows the absolute value of the angular acceleration $\alpha$ of the prominence blobs, which was highest (up to $\sim 7 \, \mu$rad\,s$^{-2}$) during the first 20 minutes of the eruptive dynamics. The acceleration of plasma blobs observed after 6:50 UT stayed below $2 \, \mu$rad\,s$^{-2}$, with many blobs showing a much smaller acceleration, as the logarithmic plot in Figure \ref{fig_dynamics}(b) indicates.

Figure \ref{fig_dynamics}(c-d) shows the estimated net magnetic torque acting on the blobs. The elevated angular acceleration during the initial stage of the eruption implies an increased torque acting on the moving plasma. Indeed, many of the blobs that were characterized by higher than average acceleration were also described by increased torque values, including the largest detected torque of $(6.3 \pm 2.3) \times 10^{23}$ N m measured soon after the beginning of the studied time interval. We also identified a group of prominence blobs with large accelerations but average torques. These prominence pieces were described by relatively small mass, allowing the magnetic force to perturb the ballistic motion without exerting a significant net torque. 

The estimated tangential magnetic force densities are plotted in Figure \ref{fig_dynamics}(e-f). As a volumetric measure, the force density does not directly depend on the mass of the moving plasma blob and should therefore closely follow the acceleration (Fig.\ \ref{fig_dynamics}(a-b)). The initial 15-20 minutes of the prominence dynamics were described by a systematically larger magnetic force, reaching $f_B \sim 3\times10^{-1}$ N m$^{-3}$ for one of the prominence blobs. Later, the force density decreased on average, although some of the blobs detected in the middle of the time interval continued to experience significant non-central forces comparable with those observed at the early stage of the eruption.

The coronal magnetic field (Fig.\ \ref{fig_dynamics}(g-h)) evaluated using Eq.\ (\ref{eq_B_field}) was the strongest (up to 135 G) for the prominence debris tracked at the beginning of the eruption. These plasma fragments were accelerated along newly reconnected magnetic field lines that linked AR 11226 and the adjacent AR 11227 \citep{driel14}. During the falling phase of the eruption (after about 6:40 UT), the estimated magnetic field of most of the blobs varied between about 5 G and 60 G. 

\begin{figure}
\begin{center}
\includegraphics[width=8.5 cm]{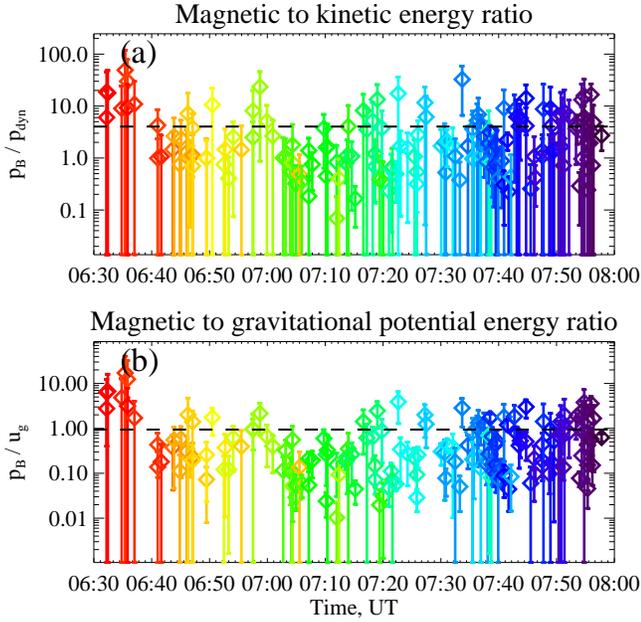}
\caption{\label{fig_pressures} Energy ratios describing the studied prominence dynamics. Diamond symbols show mean ratios for each event, vertical bars are the standard deviations reflecting the temporal spread of the ratio during the event. Dashed horizontal lines mark the ensemble-averaged ratio values. Color coding is the same as in Figure \ref{fig_events}.}
\end{center}
\end{figure}

\begin{figure}
\begin{center}
\includegraphics[width=8.5 cm]{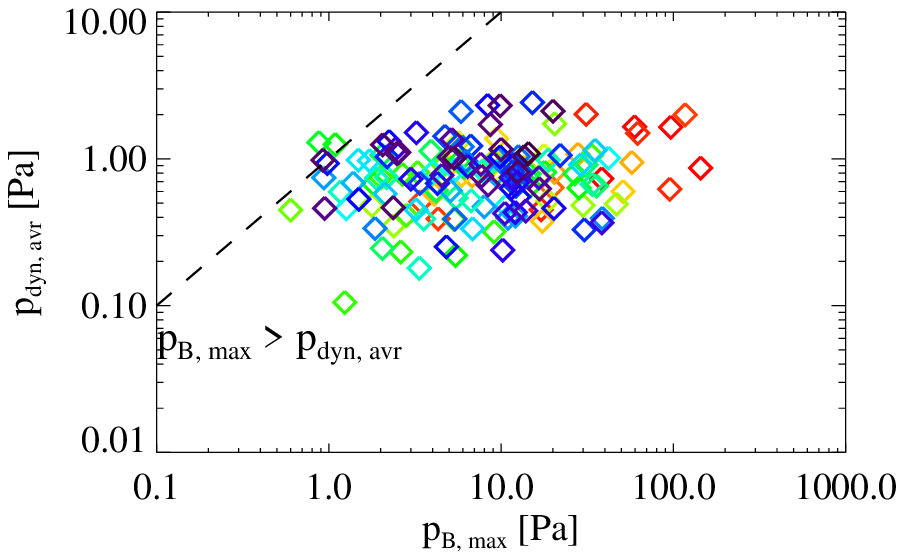}
\caption{\label{fig_pressure_scatterplot}  Average dynamic pressure exerted by a moving prominence blob versus the maximum (peak) magnetic pressure experienced by the same blob. For the events to the right of the dashed line, the instantaneous peak magnetic pressure exceeded the dynamic pressure. Color coding is the same as in Figure \ref{fig_events}. }
\end{center}
\end{figure}

Figure \ref{fig_pressures} shows two pressure/energy ratios estimated for each prominence blob that further clarify the critical role played by the coronal magnetic field in the eruptive prominence evolution. The magnetic to dynamic pressure ratio (Fig.\ \ref{fig_pressures}(a)) reflects the degree to which the magnetic force exerted by the frozen-in magnetic field threading the moving plasma can perturb its ballistic motion.  As can be seen, the value of $p_B/p_{dyn}$ exceeds 1 for most of the blobs and reaches 50 for the blobs observed at the initial stage of the eruption; the average pressure ratio (shown by the dashed horizontal line) for all the detected blobs is about 4. This implies that the force exerted by the magnetic field on the rising and falling prominence plasma was quite significant during the entire eruptive event. 

The vertical error bars on both panels of Figure \ref{fig_pressures} show the propagated statistical uncertainties (standard errors) of each measurement. The bars appear asymmetric because of the logarithmic scales used. Some of the $p_B/p_{dyn}$ uncertainties are quite large, indicating that the magnetic to dynamic pressure ratio could be much larger than the average ratios at certain locations along the blob trajectory compared to its averaged value.

The estimated magnetic to gravitational energy ratios (Fig.\ \ref{fig_pressures}(b)) are smaller than the $p_B/p_{dyn}$ ratios. It should be noted that our method of the evaluation of the gravitational potential energy relative to the nominal solar surface (see Eq.\ \ref{eq_grav_energy}) is rather inaccurate because it is based on the average, rather than actual, altitude of the falling plasma blobs. As a result, the calculation errors of the plotted $p_B/u_g$ ratios could exceed the measurement uncertainties. While the individual measurements are likely unreliable, the event-averaged $p_B/u_g$ ratio ($ \sim$ 1) can be trusted at least within a factor of two. This suggests that the solar gravity's contribution to the eruptive dynamics is of the same order as that of the magnetic field, although the latter played a prevailing role during the explosive phase of the eruption when $p_B/u_g$ systematically exceeded 1, and it was greater than 10 for several selected blobs. 

Figure \ref{fig_pressure_scatterplot} shows the time-averaged dynamic pressure $p_{dyn}$ of each plasma blob versus the peak 
value $p_{B max, i}$ of the magnetic pressure observed during the 
tracking history of the blob. The two pressure estimates are equal along the straight dashed line added to the scatter plot. The fact that nearly all blob events lie to the right of this line indicates that the instantaneous magnetic pressure enhancements experienced by a moving plasma blob are quite significant. The color coding shows that the highest maximum values of the magnetic pressure were observed at the beginning of the studied time interval, reflecting the initial acceleration of the material by the magnetic field of the prominence.

\begin{figure*}
\begin{center} 
\includegraphics[width=14 cm]{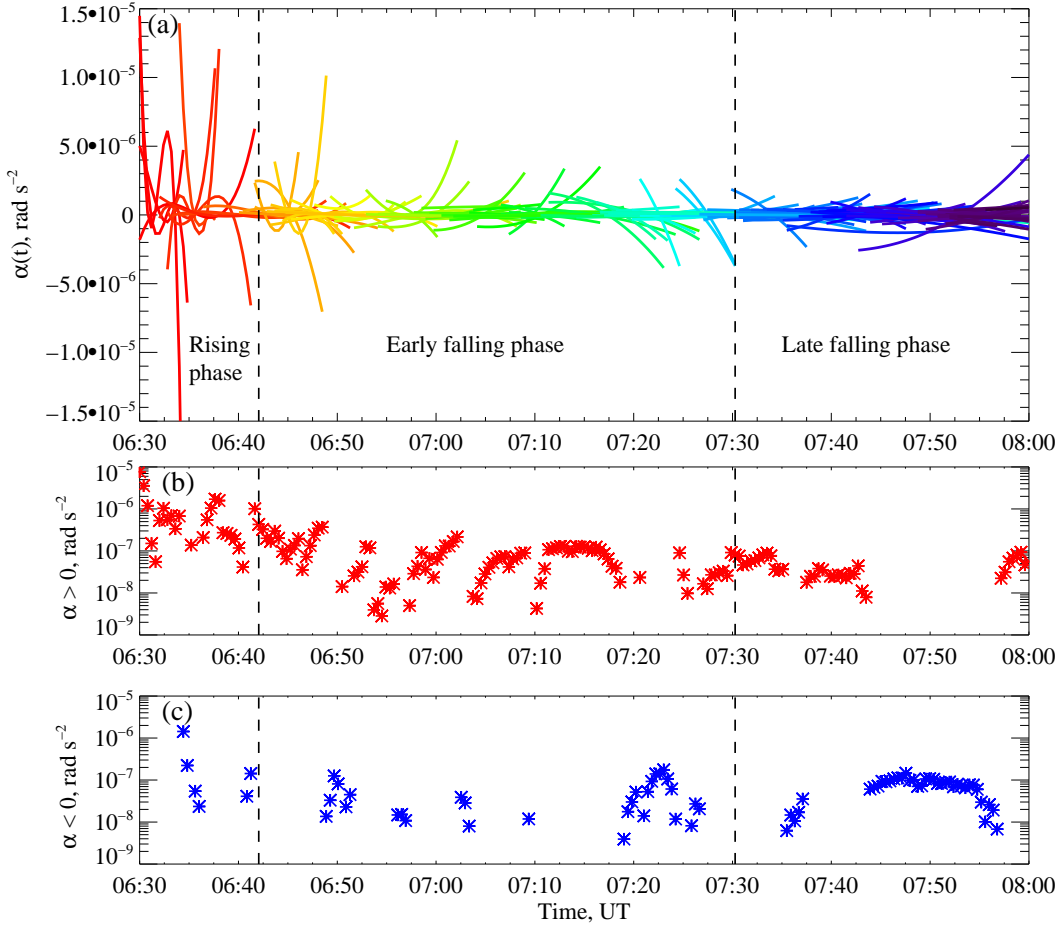}
\caption{\label{fig_accelerations} Temporal evolution of the signed instantaneous angular acceleration measured by KODA in all successfully tracked plasma blobs during the 2011 June 07 eruption. (a) Dynamics of angular acceleration describing each trajectory; color coding is the same as in Figure \ref{fig_events}. (b,c) Ensemble-averaged magnitudes of positive (b) and negative (c) acceleration values plotted on the logarithmic scale. Dashed vertical lines show the boundaries of the three main phases of the eruptive dynamics used in Tables \ref{tab:parameters} and \ref{tab:totals}. }
\end{center}
\end{figure*}

Figure \ref{fig_accelerations} provides more details on the transient dynamics of the prominence  debris. Panel (a) shows the evolution of the signed angular acceleration experienced by each plasma blob. Panels (b) and (c) show the magnitudes of the positive and negative ensemble-averaged accelerations, respectively: 
\begin{equation}
\alpha(t) = N_t^{-1}\sum_{i} \alpha_i(t),
\end{equation}
where $\alpha_i(t)$ is the time-dependent acceleration of the $i^{\text{th}}$ prominence blob and the summation is performed over all $N_t$ blobs observed simultaneously at a given time $t$. It is evident from Figure \ref{fig_accelerations} that positive acceleration prevailed during the initial stage of the studied dynamics. During this time, the magnitude of the acceleration was the largest, and the altitude of most of the prominence blobs ejected by the eruption continued to increase. This stage of the eruptive evolution, denoted on the Figure as the ``rising phase,'' lasted until about 06:42 UT. After that, the prominence plasma started to fall down, the acceleration magnitude declined, and the sign of the acceleration became negative for many blobs. We divide this behavior into the ``early falling phase'' lasting until about 7:30 UT, when positive acceleration was observed more frequently than negative acceleration, and the subsequent ``late falling phase'' which continued through the end of the studied time interval. During the late falling phase, negative angular acceleration prevailed and, with few exceptions, the acceleration magnitude was substantially smaller than that during the preceding phase. 
The time boundaries between the three eruptive phases indicated in Figure \ref{fig_accelerations} are well defined but somewhat arbitrary, and they are not intended to communicate rigorous information about physical conditions affecting the erupted prominence. Their main purpose is to assist understanding of the long-term evolution of the event. Table \ref{tab:parameters} presents the results of the quantitative analysis for the total set of plasma parameters estimated by our method over the three phases of evolution.

The first two columns of Table \ref{tab:parameters} provide the notation, physical units, and description of each computed parameter. The next three columns report the mean values and the $95\%$ confidence intervals for each parameter evaluated over the three eruption phases labeled in Figure \ref{fig_accelerations}; $n$ is the number of blobs detected during each phase. If a blob's lifetime crossed the time boundary between sequential phases, the starting time $t_{1,i}$ of the blob was used to identify the phase. The last column contains the mean values and confidence intervals for all the blobs detected during the entire prominence eruption.

\begin{deluxetable*}{ll|ccc|c}
\tablecaption{Summary of KODA measurements of the 2011 June 07 eruption event: mean values and standard errors of the estimated parameters of prominence blobs \label{tab:parameters}}
\tablewidth{0pt}
\tablehead{
\colhead{Parameter} & \colhead{Description} & \colhead{Rising phase} & \colhead{Early falling phase} & \colhead{Late falling phase} & \colhead{Whole event} \\ 
\colhead{} & \colhead{} & \colhead{($n$=11)} & \colhead{($n$=70)} & \colhead{($n$=79)} & \colhead{($n$=160) }  
}
\decimalcolnumbers
\startdata
$\alpha \,\, [\mu$rad s$^{-2}$] & Signed acceleration & 0.774 $\pm$ 0.557 & 0.077 $\pm$ 0.085 & 0.011 $\pm$ 0.055 & 0.092 $\pm$ 0.066 \\
$|\alpha| \,\,[\mu$rad s$^{-2}$] & Unsigned acceleration & 1.944 $\pm$ 1.081 & 0.495 $\pm$ 0.130 & 0.299 $\pm$ 0.059 & 0.498 $\pm$ 0.114 \\
$\tau_B $ [$\times 10^{22}$ N m] & Magnetic torque & 6.983 $\pm$ 11.038 & 0.595 $\pm$ 0.253 & 0.727 $\pm$ 0.492 & 1.099 $\pm$ 0.812 \\
$f_B \,\,[\mu$N m$^{-3}$] & Magnetic force & 0.133 $\pm$ 0.061 & 0.039 $\pm$ 0.010 & 0.022 $\pm$ 0.004 & 0.037 $\pm$ 0.008 \\
$B$ [G] & Magnetic field & 57.71 $\pm$ 25.29 & 17.66 $\pm$ 2.94 & 22.49 $\pm$ 3.04 & 22.80 $\pm$ 2.99 \\
$\rho$ [$\times 10^{-12}$ kg m$^{-3}$] & Blob density & 86.99 $\pm$ 17.03 & 94.16 $\pm$ 5.78 & 92.26 $\pm$ 5.96 & 92.73 $\pm$ 4.03 \\
$D$ [Mm] & Blob size & 6.69 $\pm$ 2.33 & 5.71 $\pm$ 0.62 & 6.27 $\pm$ 0.80 & 6.05 $\pm$ 0.50 \\
$m [\times 10^{10}$ kg] & Blob mass & 3.31 $\pm$ 3.08 & 2.08 $\pm$ 0.61 & 2.98 $\pm$ 1.02 & 2.61 $\pm$ 0.61 \\
$v$ [km s$^{-1}$] & Blob speed & 149 $\pm$ 22 & 118 $\pm$ 5 & 127 $\pm$ 6 & 124 $\pm$ 4 \\
$v_A$ [km s$^{-1}$] & Alfv\'en speed & 533 $\pm$ 201 & 162 $\pm$ 25 & 211 $\pm$ 27 & 211 $\pm$ 26 \\
$p_B$ [Pa] & Magnetic pressure & 19.82 $\pm$ 14.49 & 1.855 $\pm$ 0.668 & 2.754 $\pm$ 0.745 & 3.534 $\pm$ 1.267 \\
$p_{dyn}$ [Pa] & Dynamic pressure & 1.025 $\pm$ 0.369 & 0.668 $\pm$ 0.066 & 0.791 $\pm$ 0.104 & 0.753 $\pm$ 0.065 \\
$p_{th}$ [Pa] & Thermal pressure & 0.047 $\pm$ 0.009 & 0.051 $\pm$ 0.003  & 0.050 $\pm$ 0.003  & 0.051 $\pm$ 0.002 \\
$u_g$ [Pa] & Grav.\ energy density & 3.339 $\pm$ 0.654 & 3.615 $\pm$ 0.222 & 3.542 $\pm$ 0.229 & 3.560 $\pm$ 0.155 \\
$p_B / p_{dyn}$ & Kinetic energy ratio & 14.296 $\pm$ 8.532 & 2.727 $\pm$ 0.974 & 3.751 $\pm$ 1.076 & 4.028 $\pm$ 0.985 \\
$p_B / u_g$ & Grav.\ energy ratio & 5.074 $\pm$ 3.200 & 0.490 $\pm$ 0.163 & 0.766 $\pm$ 0.187 & 0.941 $\pm$ 0.298 \\
$p_B / p_{th}$ & Reciprocal plasma beta & 357.1 $\pm$ 225.2 & 34.52 $\pm$ 11.47 & 53.883 $\pm$ 13.14 & 66.26 $\pm$ 20.96 \\
\blue{$L$ [Mm]} & \blue{Travelled distance} & \blue{62.6 $\pm$ 14.2} & \blue{71.0 $\pm$ 10.5} & \blue{81.4 $\pm$ 9.8} & \blue{75.6 $\pm$ 6.8} \\
\enddata
\end{deluxetable*}

Table \ref{tab:parameters} shows that the rising phase of the prominence evolution was characterized by large mean values of the signed ($\sim 0.8 \pm 0.6$ $\mu$rad s$^{-2}$) and unsigned ($\sim 1.9 \pm 1.1$ $\mu$rad s$^{-2}$) angular acceleration $\alpha$, both of which are statistically significantly higher than the values estimated during the falling phases. The early falling phase exhibited a somewhat larger unsigned acceleration compared to the late falling phase, whereas the mean signed acceleration during the two falling phases was indistinguishable from zero, to within the measurement errors. This indicates that the prominence material overall experienced equal amounts of positive and negative tangential accelerations along the falling trajectories. 

The net magnetic torque$\tau_B$ characterizing the rising phase of the eruption was significantly larger than the torque estimated during the other two phases. However, due to a wide spread of torque values measured in the rising blob population, their confidence interval exceeds the mean, preventing meaningful quantitative comparisons between the phases. 

The mean force density $f_B$ measured during the rising phase ($0.13 \pm 0.06$ $\mu$N m$^{-3}$) was about three times larger than that in the early falling phase ($0.04 \pm 0.01$ $\mu$N m$^{-3}$), which in turn was larger than the value measured during the late phase ($0.02 \pm 0.004$ $\mu$N m$^{-3}$). Due to the relatively small measurement uncertainties, these differences are statistically significant.

The estimated magnetic field strength $B$ during the rising phase ($58 \pm 25$ G) was a factor of three larger than that measured during the early and late falling stages, whose average magnetic fields are indistinguishable within the uncertainty of our analysis. 
The average magnetic field strength for the entire event is $23 \pm 3$ G (see the last column in Table \ref{tab:parameters}). 

The mean mass density $\rho$ ($9\times10^{-11}$ kg m$^{-3}$), linear size $D$ ($6$ Mm), and mass $m$ ($3\times10^{10}$ kg) of the prominence blobs were found to be roughly the same during the three phases of the eruptive evolution covering the studied 1.5-hour time interval. 

The estimated blob speed $v$ also varied rather little, being marginally larger during the rising phase ($150$ km s$^{-1}$) than during the early falling phase ($120$ km s$^{-1}$), followed by a statistically insignificant increase during the late falling phase ($130$ km s$^{-1}$). All of these values are highly supersonic for the 30 km s$^{-1}$ sound speed in the cool plasma. They are slightly but significantly subAlfv\'enic according to our estimates, with the Alfv\'en Mach number ranging from a low of about 0.3 during the rising phase, when the field strength and Alfv\'en speed were elevated, to a high of about 0.7 during the early falling phase, and averaging 0.6 overall. 

Due to the large variations in the magnetic field strengths across the event, the magnetic pressure $p_B$ characterizing the rising phase ($\approx$ 20 Pa) is an order of magnitude larger than that obtained for the two other phases ($\approx$ 2-3 Pa). Despite its high variability among the blobs early in the eruptive evolution, the differences from the late averaged values of the magnetic pressure are statistically significant. 

The variations of the dynamic pressure $p_{dyn}$ across the three phases are the within statistical uncertainties of the method, with the rising phase exhibiting the highest average ($p_{dyn} \approx$ 1 Pa). 

The mean pressure/energy ratios $p_B / p_{dyn}$ and $p_B / u_g$ were substantially and significantly higher during the rising phase compared to the falling phases. The ratios averaged over the whole event were $4.0 \pm 1.0$ and $0.9 \pm 0.3$, respectively. Clearly, the magnetic pressure dominated the kinetic pressure throughout the eruptive event, whereas the magnetic and gravitational energy densities were comparable overall although the former was larger in the rising phase. 

Following the mass density, the thermal pressure $p_{th} \approx 0.05$ Pa was nearly uniform across the phases. Its value relative to the magnetic pressure $p_B/p_{th}$ (the reciprocal plasma beta) indicates that the plasma dynamics was strongly magnetically dominated throughout the evolution. 
The plasma beta was much smaller than 1 during all three eruption phases, particularly so ($\beta \approx 2.5\times10^{-3}$) during the rising phase when the inferred magnetic-field strengths are the strongest. 

\blue{
The last line in Table \ref{tab:parameters} reports the mean values of the tracking distance $L_i$ measured during the three phases of the eruption. The results show no statistically significant difference in the average tracking distance characterizing the different phases.
}

To place the studied prominence eruption in the context of observations of the associated CME, we estimated the total mass of falling material by summing over the masses of all $N$ detected plasma blobs, 
\begin{equation}
    m_{tot} = \sum_{i=1}^N m_i.
\end{equation}
We evaluated the total kinetic ($E_K$), thermal ($E_{th}$), gravitational ($E_G$), and magnetic ($E_B$) energies
by adding up the corresponding volume-integrated energy densities of all the blobs,
\begin{equation}
\begin{aligned}
    E_K &= \sum_{i=1}^N p_{dyn, i} V_i,\\
    E_{th} &= \sum_{i=1}^N p_{th, i} V_i,\\
    E_G &= \sum_{i=1}^N u_{g, i} V_i,\\
    E_B &= \sum_{i=1}^N p_{B, i} V_i.
\end{aligned}
\end{equation}
%
%
The total magnetic flux entrained in the blobs was estimated by combining the field strengths and the POS areas of the blobs,
\begin{equation}
    \Phi_{tot} = \sum_{i=1}^N B_i S_i.
\end{equation}
The obtained total values estimated for the three phases are given in Table \ref{tab:totals}. The results show that, in general, the estimated values increased during the course of the eruption. The only exception is the magnetic energy, which attains a minimum during the early falling phase. For all quantities, the late falling phase exhibits the largest figures. The low values in the rising phase principally reflect the small number of blobs ($n=11$) tracked early in the eruption, when much of the prominence material was rising above the solar limb and became invisible. During the early and late falling phases, the dark blobs are observed against the much brighter solar disk, and far greater numbers ($n=70,79$) were tracked. Consequently, the total mass and energy estimates during the fall phases are expected to be more reliable than those during the rising phase. 

\begin{deluxetable*}{l|l|ccc|c}
\label{tab:totals}
\tablecaption{Accumulated mass, energy, and flux estimates for the 2011 June 7 eruption.}
\tablewidth{0pt}
\tablehead{
\colhead{Parameter} & \colhead{Description} & \colhead{Rising phase} & \colhead{Early falling phase } & \colhead{ Late falling phase} & \colhead{CME}
}
\decimalcolnumbers
\startdata
$m_{tot} \,\, [\times 10^{11}$ kg] & Total mass & 3.6 $\pm$ 3.4 & 14.5 $\pm$ 4.3 & 23.6 $\pm$ 8.0 & 230$^1$ \\
$E_K \,\, [\times 10^{22}$ J] & Total kinetic energy & 0.5 $\pm$ 0.6 & 1.1 $\pm$ 0.4 & 1.9 $\pm$ 0.6 & 1800$^1$ \\
$E_{th} \,\, [\times 10^{22}$ J] & Total thermal energy & 0.02 $\pm$ 0.02 & 0.08 $\pm$ 0.02 & 0.13 $\pm$ 0.04 & 76$^2$ \\
$E_G \,\, [\times 10^{22}$ J] & Total gravitational energy & 1.4 $\pm$ 1.3 & 5.6 $\pm$ 1.6 & 9.0 $\pm$ 3.1 & 440$^2$ \\
$E_{mech} \,\, [\times 10^{22}$ J] & Total mechanical energy & 1.9 $\pm$ 1.9 & 6.7 $\pm$ 2.0 & 10.9 $\pm$ 3.6 & 2240$^2$ \\
$E_B \,\, [\times 10^{22}$ J] & Total magnetic energy & 4.1 $\pm$ 3.6 & 1.9 $\pm$ 0.7 & 6.9 $\pm$ 3.6 & 2700$^3$ \\
$\Phi_{tot} \,\, [\times 10^{12}$ Wb] & Total magnetic flux & 2.4 $\pm$ 1.6 & 3.7 $\pm$ 0.8 & 7.6 $\pm$ 2.3 & 28$^4$ \\
\enddata
$^1$CDAW estimates for the SOHO/LASCO CME (https://cdaw.gsfc.nasa.gov/CME\_list/index.html).

$^2$Derived from the CDAW estimate of the CME mass; see text for details. 

$^3$Prominence magnetic free energy estimate by \citet{egorov20}.

$^4$Prominence magnetic flux estimate by \citet{yardley16}.
\end{deluxetable*}

\begin{figure*}
\begin{center}
\includegraphics[width=14 cm]{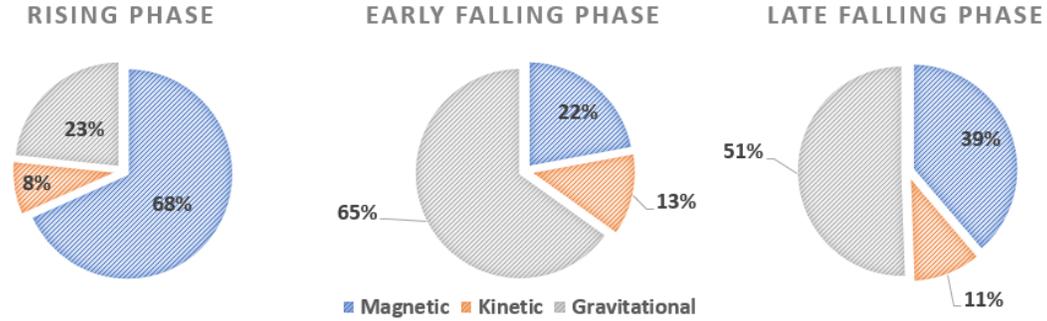}
\caption{\label{fig_energy_budget} Energy partitioning of the prominence plasma during the three eruptive phases investigated by KODA. \label{fig_budgets}}
\end{center}
\end{figure*}

Although the total estimated mass, energies, and magnetic flux in the prominence plasma were affected by the varying detection conditions above and below the limb, the partition of the energies among the different forms are expected to be more robust. The partitioning is compared across the three evolutionary phases in Figure \ref{fig_budgets}, from which we omitted the thermal energy because it is negligibly small ($<$ 1\%). The pie charts show that the magnetic energy dominated during the rising phase, fell to a minimum in the early falling phase, and then recovered significantly, though not completely, during the late falling phase. This trend tracks closely with that of the magnetic field strengths, as the blobs expand upward in the strong prominence magnetic field, reach their apexes at maximum heights where the field is weakest, and then descend to the surface at remote locations where the field strengths are intermediate between those in the prominence and in the high corona. The gravitational potential energy, which we approximated crudely by assuming an average height $h$ = 0.2$R_S$ across all phases of the eruption, was relatively small early, dominant during the early falling phase, and subsided somewhat during the late falling phase. Had it been possible to estimate the heights of the trajectories accurately, we would expect this general trend in the gravitational energy to be accentuated relative to our simple estimate, but the magnitude of its contribution would not be affected significantly. Finally, the kinetic energy was the smallest of the three energy components, and it too reached its peak percentage contribution during the early falling phase. 

For comparison with these directly measured quantities, the last column in Table \ref{tab:totals} reports some parameters of the prominence eruption deduced from CME observations and related global analyses. The mass and kinetic energy are taken directly from the Coordinated Data Analysis Workshops \citep[CDAW;][]{yashiro04} online catalogue of SOHO/LASCO CMEs at https://cdaw.gsfc.nasa.gov/CME\_list/index.html. The gravitational energy is estimated from the CME mass, 
\begin{equation}
E_{G,cme} \approx m_{cme} g_S R_S,
\end{equation} 
which is combined with the kinetic energy to yield the total mechanical energy; and the CME thermal energy likewise is estimated from the mass, 
\begin{equation}
E_{th,cme} \approx 2 \frac{m_{cme}}{m_p} k_B T_c,
\end{equation} 
using the ideal-gas law and an assumed coronal temperature $T_c$ = 2 MK. The free magnetic energy of the prominence estimated by \citet{egorov20} and the amount of magnetic flux entrained in the prominence estimated by \citet{yardley16} complete the table. We emphasize that these values correspond to the global-scale eruption, whereas our measured values pertain only to the rising and falling cool plasma blobs. The relationships between the two sets of values are discussed below. 

\begin{figure*}
\begin{center}
\includegraphics[width=17 cm]{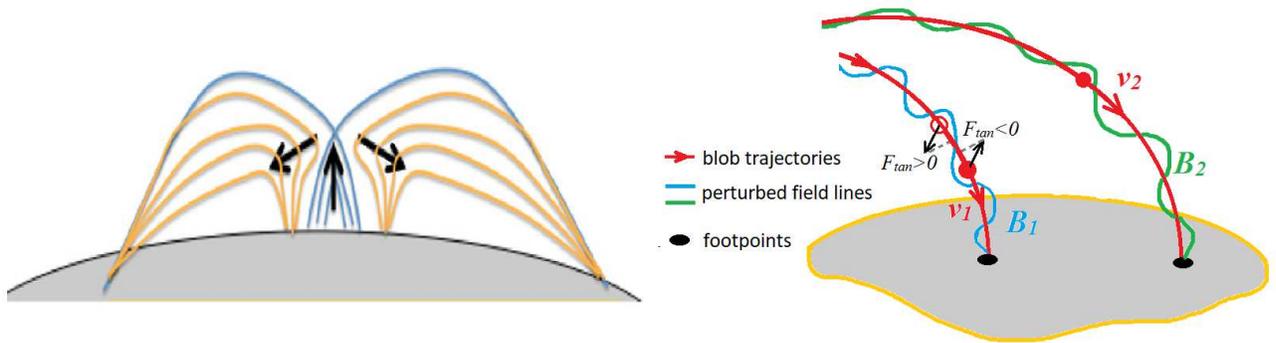}
\caption{\label{fig_mf} Left: Schematic diagram illustrating the reconfiguration of the magnetic field by reconnection during the studied eruption, according to our data-driven reconstruction. The most significant magnetic torques and forces leading to the upward and sideways accelerations (black arrows) of the plasma material are measured during the rising phase of the eruption. Right: Schematic diagram illustrating falling plasma blobs following reconnection of the magnetic field and the initial raising phase. In some cases, the motions are accompanied by small but detectable perturbations in the local magnetic forces. Distortions in the shape of the magnetic field lines (blue and green traces) at small scales, e.g.\ due to nonlinear shear Alfv\'{e}nic oscillations, can introduce alternating tangential forces perturbing the trajectories (red traces).
}
\end{center}
\end{figure*}

\section{Discussion}
\label{sec:discussion}

We applied a new remote-sensing methodology -- the Keplerian Optical Dynamics Analysis -- to dense, cool plasma blobs observed as dark features in the SDO/AIA 193 {\AA} channel during the 2011 June 7 prominence eruption, eruptive flare, and coronal mass ejection. A schematic diagram illustrating the reconfiguration of the magnetic field during this event, as suggested by our data-driven reconstruction, is shown in Figure \ref{fig_mf}. The prominence magnetic field and plasma rapidly rise from their initial quasisteady state in the low corona (vertical black arrow under blue magnetic field lines) at eruption onset. They encounter overlying and adjacent magnetic field and plasma (gold magnetic field lines) during their ascent. The ensuing magnetic reconnection redirects the prominence plasma sideways away from the eruption site (horizontal black arrows). The plasma then generally follows the magnetic field as it falls back to the solar surface, although the violent interaction also can generate nonlinear transverse Alfv\'en waves that accompany the blob motion. In three dimensions, the result is a fountain-like spray of the prominence material through a large volume of the corona. The actual magnetic configuration of the observed eruption was quite complex, as described in detail by \citet{driel14}, because the prominence encountered magnetic flux from neighboring active regions as well as from the overlying background field of the Sun. Nevertheless, the essential features of the evolution of the plasma blobs that we tracked are indicated by this simple illustrative diagram. 

KODA enabled numerous kinematic and dynamic parameters of the blobs to be deduced directly from their measured trajectories. In addition, line-of-sight integrated column masses determined by \citet{gilbert13} and temperatures determined by \citet{landi13} from multiwavelength EUV observations of the event were used to estimate the mass density, temperature, and pressure of the blobs. (\citet{carlyle14} reported the same column density as \citet{gilbert13}; \citet{landi13} found values that were higher by less than a factor of two.) Details of the results are shown in preceding figures in this paper, and several key numerical parameters are given in the tables. 

The consistency across the data set of the size of the detected, tracked blobs (6 Mm $\approx$ 9'' $\approx$ 15 AIA pixels) yields an approximately uniform mass density of about $1\times10^{-10}$ kg m$^{-3}$, equivalent to a number density of about $1\times10^{17}$ m$^{-3}$, in the prominence plasma. For the assumed temperature ($3.3\times10^4$ K), this corresponds to a prominence thermal pressure of about 0.05 Pa. The total volume of the blobs implies that a total mass of some $2\times10^{12}$ kg of cool plasma was redistributed through the corona by the eruption. Assuming that the non-ballistic motion of the blobs was due to magnetic forces acting along the extended trajectories of the blobs enabled us to estimate the strength of the magnetic field. The overall average strength was about 20 G, during the initial rising phase of the eruption the average was nearly 60 G, and some individual early blobs experienced field strengths in excess of 100 G, according to our analysis. 

The above-quoted numbers for the mass/number density, total mass, thermal pressure, and magnetic field strength in the 2011 June 7 erupting prominence are all fully consistent with accepted, measured values for active-region prominences \citep[e.g.,][]{labrosse10,mackay10,parenti14}. These prominences tend to be denser and have stronger fields, although they are smaller and more compact, compared to their counterparts in quiet Sun and the high-latitude polar crown. 

By combining the estimated field strengths with the blob sizes, we deduced that as much as $8\times10^{12}$ Wb of magnetic flux was entrained in the plasma of the observed prominence. \citet{yardley16} reported that about $3\times10^{13}$ Wb of magnetic flux cancelled at the photosphere in the vicinity of the prominence in the days prior to eruption, and that this could be a measure of the amount of flux in the prominence itself. Our results are consistent with their findings, although only within half an order of magnitude and, as should be anticipated, on the low side. Some fraction, possibly substantial, of the cancelling flux would be expected to submerge below the photosphere rather than to remain in the corona \citep[e.g.][]{harvey99}. In addition, our estimate takes into account only the fraction of the prominence flux containing cool blobs that we were able to track. It is highly unlikely that this comprises all of the magnetic flux making up the prominence structure. 

The total magnetic energy content of the blobs that we tracked ranged as high as $7\times10^{22}$ J. This is only a tiny fraction of the total magnetic energy in the prominence as a whole, however. The erupting structure was roughly 100$\times$ longer than our individual blobs \citep[600 Mm vs.\ 6 Mm; see][]{yardley16}. Hence, its total magnetic energy was almost certainly closer to $7\times10^{24}$ J, if not significantly larger still. The prominence magnetic field is not laden with a monolithic slab of cool, dense plasma; rather, it is comprised of long, thin threads and compact knots that are interspersed with hot, tenuous coronal plasma. Even this substantial amount of energy is insufficient to propel the observed eruption, however: the CME kinetic energy was estimated to be $1.8\times10^{25}$ J from SOHO/LASCO coronagraph observations. 

\citet{fainshtein16,fainshtein17} examined SDO/HMI magnetograms of the photospheric field of AR 11226 and its environs during the 2011 June 7 eruption. Subsequently, \citet{egorov20} used a nonlinear force-free model for the magnetic field configuration to attempt to reconstruct the prominence from the SDO/HMI data. They calculated a total magnetic free energy in the structure of $2.7\times10^{25}$ J and field strengths as high as 500 G in the low corona where the pre-eruption prominence was positioned. This amount of energy would be sufficient to propel the observed CME. Their field strengths are higher than our peak estimated field strengths by about a factor of four; however, our measurements were made after eruption onset, when the prominence had begun to rise, expand, and reconnect with neighboring magnetic fields. A factor of two increase in the width and height of the prominence body would reduce its peak field strength and magnetic energy content by factors of four, to about 125 G and $6\times10^{24}$ J, respectively. These numbers agree quite well with our measured field strengths and extrapolated magnetic energy. Admittedly, this close agreement may be coincidental, but it seems unlikely to be wholly accidental. 

In a completely different modeling study, \citet{petralia16} performed three-dimensional magnetohydrodynamics simulations of cool plasma blobs falling through the magnetized corona. Their goal was to replicate features of the 2011 June 7 eruption as seen in the EUV \citep{landi13,reale14}. They concluded that downfalling blobs guided by magnetic fields whose strengths are in the range 10-20 G best fit the observed motions. Their values are in excellent agreement with our estimates during the falling phases of the eruption, which we obtained by directly analyzing the blob trajectories. 

The key inferences that we draw from this discussion are that our findings are quantitatively consistent with other measures of the 2011 June 7 event in particular, and of active-region prominence eruptions in general. All of those alternative measures were derived using very different data sets and techniques, and by numerous investigators. 

\section{Conclusions}
\label{sec:conclusions}

We presented a new approach to deducing remotely sensed physical parameters of solar corona based on an analysis of the trajectories of rising and falling prominence plasma. The KODA methodology involves spatiotemporal tracking of prominence debris and a quantitative evaluation of the perturbations to their ballistic trajectories caused by the coronal magnetic field. 

The conducted analysis allowed us, for the first time, to measure physical characteristics of the coronal plasma across a vast spatial domain surrounding the eruption. The following main conclusions were reached by applying KODA to the 2011 June 7 prominence eruption:

(1) The coronal magnetic field introduces substantial perturbations to the projectile motions of most of the studied prominence blobs. The detected effect is statistically significant for the detected population of 160 individual plasma blobs, and is especially evident for the blobs whose trajectories were tracked over distances longer than 50 Mm. 

(2) The magnetic forces perturbing the trajectories of the falling plasma blobs can manifest themselves in both positive and negative tangential acceleration along the same trajectory. The oscillatory character of the perturbations in these cases suggests a scenario in which the falling plasma undulates in response to transverse waves on the affected coronal loops, in addition to secular magnetic tension forces that reflect a large-scale guiding of the falling prominence material by the newly reconnected loops.

(3) The estimated magnitude of the coronal magnetic field ($\sim$ 20-60 G) is consistent with well established direct measurements for prominences in active regions. The volumetric density of the magnetic force acting on the prominence plasma is estimated to be in the range $0.022$-$0.133$ $\mu$N m$^{-3}$. The magnetic field and the associated magnetic force are found to be strongest during the initial phase of the eruptive dynamics.

(4) The ratio of the magnetic pressure exerted by the coronal field to the dynamic pressure describing moving plasma blobs was found to be about 4 on average but reaches 14 for the blobs observed shortly after the eruption, suggesting that magnetic forces accelerated the plasma over the course of the eruption.

(5) The total mass ($\sim 10^{12}$ kg) and the total kinetic, gravitational, and magnetic energies (all $\sim 10^{22}$ J) of the studied plasma blobs are significant, although they are substantially smaller than the total mass ($\sim 2\times10^{13}$ kg), kinetic energy ($\sim 2\times10^{25}$ J), and gravitational energy ($\sim 4\times10^{24}$ J) of the associated halo CME.

(6) The magnetic energy contained in the prominence blobs dominated the kinetic energy throughout, and the gravitational energy during the initial phase of the eruptive evolution, but was smaller than the gravitational energy during the falling phase of evolution when the dynamics of the plasma blobs were closer to ballistic.

Direct measurements of prominences during the quasistatic phases of their lifetimes provide critical insight into their nature, structure, and evolution. Our novel indirect measurement technique, applied to a highly dynamic prominence eruption, complements the well-established spectroscopic methods. The application of KODA to other well-observed solar eruptive events is planned and, we anticipate, will yield additional new and valuable insights into their behavior and characteristics. 

\acknowledgements

V.M.U.\ gratefully acknowledges Dr.\ Yury Muzalevsky (Muzalevskii) for useful discussions and advice on data analysis techniques. V.M.U. was supported through the Partnership for Heliophysics and Space Environment Research (NASA grant 80NSSC21M0180); B.J.T.\ and C.R.D.\ were supported by NASA H-ISFM and predecessor H-GI and H-SR grants at Goddard Space Flight Center. The CDAW SOHO/LASCO CME catalog is generated and maintained by The Catholic University of America and NASA Goddard in cooperation with the U.S.\ Naval Research Laboratory. SOHO is a project of international cooperation between ESA and NASA. 

\bibliography{references}{}
\bibliographystyle{aasjournal}

\end{document}